\documentclass[a4paper,11pt]{article}

\usepackage{jheppub}
\usepackage{graphicx,tabularx}
\usepackage{amsmath, amssymb, amsthm, amsfonts}
\usepackage[dvipsnames]{xcolor}
\usepackage{soul}
\usepackage{comment}
\usepackage{slashed}
\usepackage[yyyymmdd,hhmmss]{datetime}
\usepackage[backgroundcolor=white, bordercolor=white, textsize=tiny]{todonotes}

\providecommand{\tightlist}{%
\setlength{\itemsep}{1pt}\setlength{\parskip}{1pt}}

\newcommand{\tev}{\text{TeV}}
\newcommand{\gev}{\text{GeV}}
\newcommand{\ifb}{\text{fb}^{-1}}
\newcommand{\iab}{\text{ab}^{-1}}
\newcommand{\nb}{\text{nb}}
\newcommand{\pb}{\text{pb}}
\newcommand{\fb}{\text{fb}}

\newcommand{\sbeta}{s_\beta}
\newcommand{\cbeta}{c_\beta}
\newcommand{\tbeta}{t_\beta}

\newcommand{\cba}{c_{\beta-\alpha}}

\newcommand{\gfive}{\gamma^5}
\newcommand{\ctbeta}{t_\beta^{-1}}

\newcommand{\mh}{m_h}
\newcommand{\mH}{m_H}
\newcommand{\mA}{m_A}
\newcommand{\mC}{m_{H^\pm}}
\newcommand{\mx}{m_{12}^2}
\newcommand{\met}{\slash{\hspace{-2.5mm}E}_T}

\newcommand{\bpa}{\textbf{BP-A}}
\newcommand{\bpb}{\textbf{BP-B}}

\renewcommand{\eqref}[1]{Eq.~(\ref{#1})}
\newcommand{\secref}[1]{Sec.~\ref{sec:#1}}
\newcommand{\figref}[1]{Fig.~\ref{fig:#1}}

\newcommand{\appref}[1]{Appendix~\ref{sec:#1}}

\newcommand{\be}{\begin{equation}\begin{aligned}}
\newcommand{\ee}{\end{aligned}\end{equation}}

\title{Exotic Higgs Decays in Type-II 2HDMs at the LHC
and Future 100 TeV Hadron Colliders}
\author[a]{Felix Kling,}
\author[b]{Honglei Li,}
\author[c]{Adarsh Pyarelal,}
\author[d]{Huayang Song,}
\author[d]{Shufang Su}

\affiliation[a]{
  Department of Physics and Astronomy,
  University of California,
  Irvine, CA 92697, USA}
  
\affiliation[b]{
  School of Physics and Technology,
  University of Jinan,
  Jinan Shandong 250022,
  China
}

\affiliation[c]{
  School of Information,
  University of Arizona,
  Tucson, Arizona 85721, USA
}

\affiliation[d]{
  Department of Physics,
  University of Arizona,
  Tucson, Arizona 85721, USA
}

\emailAdd{fkling@uci.edu}
\emailAdd{sps\_lihl@ujn.edu.cn}
\emailAdd{adarsh@email.arizona.edu}
\emailAdd{huayangs@email.arizona.edu}
\emailAdd{shufang@email.arizona.edu}

\preprint{
\begin{flushright}
  UCI-TR-2018-09 \\
\end{flushright}
}

\abstract{
The exotic decay modes of non-Standard Model (SM) Higgses in models with extended Higgs sectors  have the potential to serve as powerful search channels to explore the space of Two-Higgs Doublet Models (2HDMs) that cannot be studied effectively using conventional decay channels. Once kinematically allowed, heavy Higgses could decay into  pairs of light non-SM Higgses, or a non-SM Higgs and  a SM gauge boson, with  branching fractions that dominate those of the conventional decay modes to SM particles. In this study, we focus on the prospects of probing exotic decay channels at the LHC and a future 100 TeV \emph{pp} collider in the context of Type-II 2HDMs. We study the three prominent exotic decay channels, $A\rightarrow HZ$, $A\rightarrow H^\pm W^\mp$ and $H^\pm\rightarrow H W^\pm$, and find that a 100-TeV \emph{pp} collider can probe the entire region of the Type-II 2HDM parameter space that survives current theoretical and experimental constraints with exotic decay branching fraction $\gtrsim 20\%$.  
}

\begin{document} 

\titlepage
\maketitle
\newpage

\section{Introduction}
\label{sec:intro}

With the discovery of a light Standard Model (SM)-like Higgs boson at
the LHC~\cite{Aad:2012tfa,Chatrchyan:2012xdj}, the search for new physics
beyond the SM has become even more pressing, given the need to stabilize
the  the mass of the Higgs boson against large radiative
corrections.  Many of the new physics models constructed to augment the SM contain
an extended Higgs sector that is responsible for  electroweak symmetry breaking.
One of the most straightforward and well-motivated class of extensions to the SM
is the category of models collectively known as Two-Higgs-Doublet Models (2HDMs)~\cite{Branco:2011iw}.
After electroweak symmetry breaking, the spectra of  2HDMs contain five mass eigenstates
(\textit{h}, \textit{H}, \textit{A}, $H^\pm$), with the CP-even Higgs \textit{h}
being the observed SM-like Higgs. These new Higgs bosons can be  constrained 
through either indirect searches via precision measurements of Higgs
properties at future Higgs factories~\cite{Gu:2017ckc,Chen:2018shg} or direct 
searches at particle colliders at the energy frontier.
In this paper, we focus on the potential for direct discovery of these heavy
states at the Large Hadron Collider (LHC) as well as a proposed 100 TeV
$pp$ collider~\cite{CEPC-SPPCStudyGroup:2015csa,fccplan}.  

The conventional searches for neutral heavy Higgses (\emph{A} and \emph{H}) in 2HDMs 
mainly focus on  modes in which they decay into pairs 
of SM particles.
While these modes have been proven to be effective in the
search for the SM Higgs, they suffer from certain limitations 
when it comes to searches for non-SM heavy Higgses.
In  particular, current data  indicates
that the observed 125 GeV Higgs is very SM-like, which implies that the couplings
of \emph{A} and \emph{H} to the 
SM gauge bosons (\emph{W}, \emph{Z})
 are suppressed, which in turn implies the suppression
of both their production via weak-boson fusion and weak-boson associated processes,
as well as their decays to SM gauge boson pairs (\emph{WW/ZZ}).
The decay channel to a pair of top quarks, which becomes kinematically accessible
for large Higgs masses, suffers from both the large top-quark pair production 
background in the SM as well as non-trivial interference effects \cite{Carena:2016npr},
significantly reducing its sensitivity.

If the BSM Higgs sector is \textit{hierarchical} - that is, its states
are sufficiently well-separated in mass - additional decay channels open up,
for example, the decay of a heavy Higgs to two lighter Higgses, or to a lighter
Higgs and an SM gauge boson. Given the corresponding unsuppressed
couplings and the large amount of available phase space, these decay modes
can be dominant in large regions of parameter space. In this scenario, the
branching fractions of the conventional decay modes are reduced and the 
experimental search limits obtained using them are relaxed correspondingly.

The \textit{exotic} decay modes of heavy neutral Higgses to lighter Higgses,
namely $H/A\rightarrow AZ/HZ/H^\pm W^\mp$ and $H\rightarrow AA/hh/H^+H^-$,
offer alternative avenues for discovering heavy Higgses that complement 
the conventional ones. The reach of individual channels at the LHC have 
been studied in the literature~\cite{Coleppa:2014hxa,Li:2015lra}
and searches for the most promising channel, $H/A\rightarrow AZ/HZ$, have been 
carried out at both ATLAS \cite{Aaboud:2018eoy} and CMS \cite{Khachatryan:2016are}.
The current experimental data excludes heavy neutral Higgses with masses
up to about 700 -- 800 GeV, depending on the BSM Higgs spectrum and values of $\tan\beta$.
Additionally, the $A\rightarrow hZ$, $H\rightarrow hh$ channels
have also been studied at the LHC \cite{Aaboud:2017cxo,Aaboud:2018ftw,Sirunyan:2018iwt}.
However, no constraints on the 2HDM parameter space can be imposed using these 
channels, the observed Higgs boson is SM-like, corresponding to the 
alignment limit in 2HDMs, in which such channels are highly suppressed.

Charged Higgs bosons pose a special challenge for experimental searches 
\cite{Aaboud:2018cwk,Aaboud:2018gjj,CMS-PAS-HIG-16-031}.
They are dominantly produced in association with top quarks ($tbH^\pm$), with a 
cross section much smaller than that of the dominant production channels for the
neutral Higgses(gluon fusion and \emph{bbH/A} associated production   at large
$\tan\beta$).  The branching fractions of the conventional search channels,
$H^\pm \rightarrow \tau\nu, cs$ are suppressed once the decay mode
$H^\pm \rightarrow tb$ opens up.  Despite its large branching fraction,
the $H^\pm \rightarrow tb$ decay mode holds little promise
for discovering charged Higgses at the LHC due the large SM backgrounds to this process. The exotic decay modes $H^\pm \rightarrow AW^\pm/H W^\pm$ could potentially be
useful in charged Higgs searches~\cite{Coleppa:2014cca,Kling:2015uba}. 
However, the complicated decay final states and relatively large SM backgrounds  limit their reaches at the LHC.

The study in Ref.~\cite{Kling:2016opi} constructs benchmark planes for these 
exotic decay channels at the LHC, taking into account both theoretical
constraints such as perturbativity, unitarity, and vacuum stability, as
well as current experimental limits from direct and indirect searches on
the parameter space of Type-II 2HDMs. Sizable mass splittings between Higgses, required for the exotic decay modes, can be achieved for heavy Higgs masses up to about 2 TeV. 
Thus, in this paper, we focus on a subset of the benchmark scenarios in 
Ref.~\cite{Kling:2016opi} that permit TeV-scale masses, and construct 
two benchmark planes: \bpa~($m_{A}>m_{H}=m_{H^\pm}$) with $A\rightarrow HZ/H^\pm W^\mp$
and \bpb~($m_{A}=m_{H^\pm}>m_{H}$) with $A\rightarrow HZ$, $H^\pm\rightarrow H W^\pm$.

In recent years, a possible 100 TeV $pp$ collider has been discussed
worldwide,  with the two leading proposals being
the Future Circular Collider (FCC) at CERN \cite{fccplan} and the Super 
proton-proton Collider (SppC) in China \cite{CEPC-SPPCStudyGroup:2015csa}.
It is important to explore the discovery potential for new 
physics models at such a machine to establish the physics case for building it.
One advantage of such a high energy machine is that   top quarks produced 
in heavy particle decays will be highly boosted, 
resulting in fat jets
that can be effectively identified using top-tagging techniques~\cite{Plehn:2010st,Plehn:2011sj,Kling:2012up,Kaplan:2008ie,Thaler:2011gf,Kasieczka:2017nvn}.
This will allow us to distinguish new physics signals with top quarks in the
final states from the large SM backgrounds involving top quarks, which typically
pose a formidable challenge at the LHC.
 
In this paper, we study the discovery potential of   non-SM heavy Higgses 
in Type-II 2HDMs at the LHC, the High Luminosity LHC (HL-LHC), as well as 
a 100 TeV \emph{pp}  collider: 
\be
\textbf{LHC:}~\mathcal{L}=300~\ifb, \quad 
\textbf{HL-LHC:}~\mathcal{L}=3~\iab, \quad
\textbf{100 TeV:}~\mathcal{L}=3~\iab, \quad
\ee
combining all the viable exotic decay channels.
We perform a detailed collider analysis to obtain the 95\% C.L. exclusion 
limits  as well as 5$\sigma$ discovery reach for benchmark planes \bpa~and \bpb.
In recent years, multivariate analysis techniques  such as neural 
networks \cite{Aad:2012tfa}, boosted decision trees (BDT) \cite{Chatrchyan:2012xdj}, 
the Matrix Element Method \cite{ Kondo:1988yd,Gainer:2013iya} and
Information Geometry \cite{Brehmer:2016nyr,Brehmer:2017lrt} have begun to be more widely used 
in experimental particle physics searches. In our study, we construct a set 
of physics-motivated variables that we use as input features for gradient 
BDT classifiers.

The rest of the paper is organized as follows.  In \secref{2hdm},
we present a brief review of hierarchical 2HDMs and introduce the
benchmark planes \bpa~and \bpb. In \secref{HAZ}, we study the `golden' channels
$A/H\rightarrow HZ/AZ$ and explore their reach at  the LHC, HL-LHC, as well 
as a 100 TeV $pp$ collider.  In particular, we studied both the $bb\ell\ell$
and  $\tau\tau\ell\ell$ states as well as the $tt\ell\ell$ final state
using top tagging techniques to identify boosted top quarks in the final state.
In \secref{HCW}, we present the $H\rightarrow H^\pm W^\mp$ channel.
In \secref{CHW}, we explore the discovery potential for charged Higgses via the 
$H^\pm \rightarrow HW^\pm$ channel.  In \secref{reach}, we  present the combined reach in 2HDM parameter space obtained with 
these channels at the LHC and a future 100 TeV $pp$ collider. In 
\secref{conclusion}, we conclude.  \appref{method} and \appref{toptagging}
describe the methodology   used for our 
collider analysis  and how we simulate top tagging, respectively.

\section{Hierarchical Two Higgs Doublet Models: A Review}
\label{sec:2hdm}

\subsection{Properties of 2HDMs}

In this section, we provide a brief review of the aspects of 2HDMs that
are most relevant to this study. For a pedagogical introduction to this
topic, see~\cite{Kling:2016yls,Pyarelal:2017fey}. The scalar sector of 
2HDMs consists of two SU(2) doublets $\Phi_i$, with $i=1,2$, which can be
explicitly parameterized in terms of their real and complex components
as shown below.
\be
  \Phi_i =
  \begin{pmatrix}
    \phi_i^+\\
    (v_i+\phi_i+i\varphi_i)/\sqrt{2}
  \end{pmatrix}
\ee

\noindent Here, $v_i$ are the vacuum expectation values (VEVs) for the
neutral components of the doublets, satisfying the condition $v_1^2 +
v_2^2 = v^2$, with $v=246~\gev$.  This allows us to introduce the mixing
angle $\beta$ such that $\tan\beta=v_2/v_1$\footnote{In this paper we often
employ the shorthand notation $s_\theta, c_\theta,
  t_\theta=\sin \theta, \cos\theta, \tan\theta$.}%
. Assuming CP conservation and a softly-broken $\mathcal{Z}_2$ symmetry\footnote{The
most general scalar potential also contains the term
  $\left[ \lambda_6(\Phi^\dagger_1 \Phi_1)+\lambda_7 (\Phi^\dagger_2
  \Phi_2) \right](\Phi^\dagger_1 \Phi_2)+h.c.$ and potentially leads to
  flavor-changing neutral currents (FCNC). In the following we will
  neglect this term by imposing a $\mathcal{Z}_2$ symmetry under which the
scalar fields transform as $\Phi_1 \to -\Phi_1$ and $\Phi_2 \to
\Phi_2$.}, the scalar portion of the 2HDM Lagrangian can be written
down as
\begin{align}
  \newcommand{\Phisq}[2]{\Phi^\dagger_{#1}\Phi_{#2}}
  \newcommand{\massterm}[4]{m_{#1 #2}^2\Phisq{#3}{#4}}
  \begin{split}
    V(\Phi_1, \Phi_2)
    &=  \massterm{1}{1}{1}{1} + \massterm{2}{2}{2}{2}
    - m^2 _{12}(\Phisq{1}{2} + h.c.)
    + \frac{\lambda_1}{2} (\Phisq{1}{1})^2
    + \frac{\lambda_2}{2} (\Phisq{2}{2})^2\\
    &+ \lambda_3 (\Phisq{1}{1}) (\Phisq{2}{2})
    + \lambda_4 (\Phisq{1}{2})(\Phisq{2}{1})
    + \frac{1}{2} \left[\lambda_5(\Phisq{1}{2})^2 + h.c.\right].
  \end{split}
  \label{eq:potential}
\end{align}

After the mechanism of electroweak symmetry breaking (EWSB), the scalar
sector of a 2HDM consists of five mass eigenstates: a pair of neutral
CP-even Higgses, \emph{h} and \emph{H}, a CP-odd Higgs, \emph{A}, and a
pair of charged Higgses $H^\pm$. For these states we can write
\be \centering 
h &= - s_\alpha  \, \phi_1 + c_\alpha  \, \phi_2, 
\quad\quad\quad &A =& - s_\beta \, \varphi_1 \, + c_\beta \, \varphi_2, 
\\ H &=  \phantom{-} c_\alpha \, \phi_1 + s_\alpha \, \phi_2, 
\quad\quad\quad &H^\pm =& - s_\beta \, \phi_1^\pm + c_\beta\,
\phi_2^\pm. 
\ee

\noindent In the following, we will identify \emph{h} with the
discovered SM-like 125 GeV Higgs~\footnote{This is slightly different from the usual convention that the mass eigenstates $h^0$ and $H^0$ are ordered by their masses.  In this study, $h$ can either be the light one or the heavy one.  In our discussion of the collider study below, which focusses on heavy BSM Higgs boson, $H$ is typically taken to be the heavy CP-even Higgs, although $H$ being the light CP-even Higgs is still a viable possibility given the current experimental search results~\cite{Coleppa:2013dya}.} and study the collider reach
of heavy non-SM Higgses. 

The potential in Eq.~(\ref{eq:potential}) contains eight 
independent parameters: three mass parameters  $m^2_{11, 22, 12}$ and five 
quartic couplings $\lambda_{1, 2, 3, 4, 5}$. For our purposes, it is convenient to
parameterize  2HDMs  by the physical Higgs masses, $m_h$, $m_H$, $m_A$ and 
$m_{H^\pm}$, the mixing angle between the two CP-even Higgses
$\alpha$, $\tan\beta$, the electroweak VEV $v$, and the soft 
$\mathcal{Z}_2$ symmetry breaking parameter $\mx$.
Two of these parameters, namely the vacuum expectation value $v$ and
the mass of the SM-like Higgs, $m_h$ are known to be 246 GeV and 125 GeV
respectively, leaving the remaining six independent parameters.
Note that in a generic 2HDM, there
are no mass relations between the Higgs states, and therefore exotic
Higgs decays such as $A \to HZ$ are possible.

As mentioned earlier, we have introduced a $\mathcal{Z}_2$ symmetry to
avoid tree-level FCNCs, which implies that each fermion type is only
allowed to couple to one Higgs doublet. In this work we will focus on
Type-II 2HDM, in which the up-type quarks only couple to $\Phi_2$, and the
down-type quarks and leptons only couple to $\Phi_1$.


\subsection{Couplings in the Alignment Limit}

The most recent data from the LHC indicate that the coupling strength of
the recently discovered 125 GeV Higgs boson is consistent with the
SM~\cite{Khachatryan:2016vau}. In the context of a 2HDM, this can
naturally be achieved in the \textit{alignment limit}, where $\cba=0$, with \emph{h} being
identified with the SM Higgs in our convention. Its couplings to fermions and gauge
bosons are precisely those predicted by the SM.

Any deviation of the signal strength of the SM-like Higgs $h$  from its SM
prediction will constitute clear evidence for new physics and provide
strong motivation for additional experimental searches to understand its
nature. In the absence of such deviations at the LHC, or possibly a
future lepton collider, future limits will further push us towards the
alignment limit~\cite{Coleppa:2014hxa, Gu:2017ckc,Chen:2018shg}. For this 
reason, the following discussion will assume $\cba=0$.  A discussion of the 
more general case can be found in~\cite{Kling:2016opi}.

Near the alignment limit, the coupling of the SM-like Higgs \emph{h} to pairs
of gauge bosons $V = Z, W^\pm$ is SM-like, while the coupling of the
heavier CP-even neutral Higgs $H$ to gauge boson pairs is suppressed,
$g_{HVV} \sim \cba$. Furthermore, the couplings of $h$
to a heavier scalar and a gauge boson $g_{hAZ}\sim g_{h H^\pm W^\mp} \sim \cba$
are also suppressed. The unsuppressed\footnote{Note that the couplings of two CP-even or CP-odd Higgses 
  to the \emph{Z}-boson, as well as the coupling of two $Z$-bosons and a 
  CP-odd Higgs, vanish since such a coupling would violate CP-invariance. 
  A coupling of the charged scalar $H^\pm$ to a pair of vector bosons at 
  most appears at loop level.} 
  couplings of the additional scalars to vector bosons in the alignment limit are
given by
\be
\!\!\!
g_{HAZ} 			\!=\! \frac{m_Z}{v}  (p_H^\mu\!-\!p_A^\mu), \ \ \
g_{HH^\pm W^\mp} 	\!=\! \pm \frac{im_W}{v}  (p_H^\mu\!-\!p_{H^\pm}^\mu), \ \ \
g_{AH^\pm W^\mp}    \!=\! \frac{m_W}{v}  (p_{H^\pm}^\mu \!\! - \!p_A^\mu), 
\ee
where $p^\mu_X$ represents the outgoing momentum for particle
\emph{X}. We can see that the non-SM like Higgses   have unsuppressed 
couplings only to the other non-SM like Higgses, but suppressed couplings to 
the SM-like Higgs and pairs of gauge bosons. Therefore, only the heavier 
non-SM Higgs will decay into a
lighter non-SM like Higgs and a gauge boson via an exotic decay mode.
The lightest non-SM like Higgs will then decay into fermion pairs. In the Type-II 2HDM, the couplings of the non-SM Higgses to SM fermion pairs in the alignment limit can be written as
\be
\!\!\! g_{Huu} 	= - g_{Auu} \gfive 	=   y_u  \ctbeta  	,\quad
g_{Hdd} 	 	=   g_{Add}  \gfive	= - y_d \tbeta 	,\quad
g_{H\ell\ell} 	=   g_{A\ell\ell}\gfive 	= - y_\ell \tbeta,
\ee
where $y_f$ are the SM  fermion Yukawa couplings. Note that
the fermion coupling for both heavy neutral scalars, $A$ and $H$, have
the same scaling with the mixing angle $\beta$ under the alignment
limit. The couplings of the charged Higgs boson to the fermions are
\be
g_{H^\pm u_id_j}	\!=\! \frac{V_{ij}}{\sqrt{2} } \left[ (t_\beta y_d \!+\! t_\beta^{-1} y_u)
			\!+\! (t_\beta y_d \!-\! t_\beta^{-1} y_u) \gfive \right], \quad\quad 
 g_{H^\pm \ell \nu}	\!=\! \frac{ t_\beta y_\ell }{\sqrt{2} } (1\!+\! \gfive) \ .
\ee
%


\subsection{Constraints on Hierarchical 2HDMs}
\label{sec:constraints}

To understand the theoretical constraints on 2HDMs, it is useful to
consider the relations between the quartic couplings and the physical
masses. In the alignment limit, we can express the quartic couplings of
the scalar potential as follows \cite{Kling:2016opi}.
\be
v^2 \lambda_1 &= m_h^2 - \tbeta^{2\phantom{-}} \left[\frac{m_{12}^2}{\sbeta\cbeta}-m_H^2 \right] ,
\quad\quad
v^2 \lambda_4  = m_A^2 - 2 m_{H^\pm}^2 + m_H^2 \,
+  \left[\frac{m_{12}^2}{\sbeta\cbeta}-m_H^2 \right] ,\\
v^2 \lambda_2 &= m_h^2 - \tbeta^{-2} \left[\frac{m_{12}^2}{\sbeta\cbeta}-m_H^2 \right] ,
\quad\quad
v^2 \lambda_5 = m_H^2 - m_A^2 + \left[\frac{m_{12}^2}{\sbeta\cbeta}-m_H^2 \right] ,\\
 v^2 \lambda_3 &= m_h^2 + 2 m_{H^\pm}^2 - 2m_H^2
- \left[\frac{m_{12}^2}{\sbeta\cbeta}-m_H^2 \right] .
\ee
We can see that the soft $\mathcal{Z}_2$ breaking term
$m_{12}^2$ plays a crucial role, as  it affects the size of the  
trilinear and quartic scalar self-couplings. As discussed
in~\cite{Kling:2016opi}, its possible allowed values are dictated by
requiring vacuum stability and tree-level unitarity of the theory. The
latter roughly requires the quartic couplings to be perturbative,
$\lambda_i \lesssim 4\pi$. Thus, perturbativity of $\lambda_{1,2}$
requires $|\mx -\mH^2 \sbeta \cbeta| \lesssim v^2$, which naturally
leads us to fix the coefficient of the soft $\mathcal{Z}_2$ breaking
term in the Lagrangian to be
\be
m_{12}^2 = m_H^2 \sbeta \cbeta.
\label{eq:fixm122}
\ee
It is possible to deviate from this relation for values of
$\tbeta$ close to unity and for low scalar masses $\mH \sim v$. However,
in this study we focus on the high scalar mass region that can be probed
at a future high energy collider and we therefore require
\eqref{eq:fixm122} to hold for the rest of the paper.

In the following, we summarize the theoretical and experimental constraints
on the 2HDM parameter space, and their implications for exotic Higgs
decays. We only consider the alignment limit $\cba=0$ and require
$m_{12}^2 = m_H^2 \sbeta \cbeta$. A more detailed discussion is
presented in~\cite{Kling:2016opi}.

\begin{description}
    
\item[Vacuum Stability]

In order to have a stable electroweak vacuum~\cite{Gunion:2002zf}, the
following scalar mass conditions need to be fulfilled:
\be
\mh^2+\mC^2 -\mH^2 >0 , \quad \text{and}\quad \mh^2+\mA^2-\mH^2 >0 \, .
\label{eq:stability}
\ee
This implies that for $m_H > m_{A, H^\pm}$,  the mass
splittings between the heavy CP-even Higgs \emph{H} and the other heavy
scalars \emph{A} and $H^{\pm}$ have to be small, such that the decays of $H$
into the \emph{AZ, AA,} $H^+ H^{-}$ and $H^\pm W^{\mp}$ final states 
are not kinematically allowed.

\item[Tree-Level Unitarity]
Requiring tree-level unitarity of the scattering matrix in the 2HDM scalar
sector~\cite{Ginzburg:2005dt} imposes the following additional mass constraints:
\be
\!\!\!\!
|  \mH^2 \!-\!   \mA^2                    | \!<\! 8\pi v^2  , \quad
|3\mH^2 \!+\!  \mA^2 \!-\!  4\mC^2| \!<\! 8\pi v^2  , \quad
|  \mH^2 \!+\!  \mA^2 \!-\!  2\mC^2| \!<\! 8\pi v^2   , \\
|3\mH^2 \!-\!   \mA^2 \!-\!  2\mC^2| \!<\! 8\pi v^2  , \quad
|3\mH^2 \!-\! 5\mA^2 \!+\! 2\mC^2| \!<\! 8\pi v^2  .  \quad\quad\quad\quad
\ee
Here we have ignored sub-leading terms proportional to $\mh^2$.
Note that these constraints are independent of the value of $\tbeta$.

\item[Electroweak Precision Measurements]

Measurements of electroweak precision observables impose strong constraints on
the 2HDM mass spectrum~\cite{Haller:2018nnx}. In particular, these constraints require the
charged scalar mass to be close to the mass of one of the heavy neutral
scalars.
\be
\mC \approx \mH \quad \text{or} \quad \mC\approx \mA.
\ee

\item[Flavour Constraints]

Various flavor measurements~\cite{Amhis:2016xyh, Haller:2018nnx} provide indirect
constraints on the 2HDM parameter space, in particular on the mass of
the charged scalar. The most stringent of these comes from the
measurement of the branching fraction for the decays $b \to s \gamma$
and $B^+ \to\tau \nu$, which disfavor $\mC < 580~\gev$
\cite{Misiak:2017bgg} and large values of $\tbeta$ respectively.  
Flavor constraints, however, can be alleviated with contributions from other 
sectors of new physics models~\cite{Han:2013mga}.
In this paper, we focus on the direct collider reach 
of heavy Higgses without imposing the flavor constraints.

\item[Direct Searches at LEP and LHC]

While the search for pair-produced charged Higgs bosons at the Large
Electron-Positron Collider (LEP) imposes a lower bound of 80 GeV on the
mass of the charged Higgs boson~\cite{Abbiendi:2013hk}, LEP searches for 
$AH$ production constrain the sum of the masses $m_H + m_A > 209~\gev$~\cite{Schael:2006cr}.
LEP bounds on single neutral Higgs production do not apply 
in the alignment limit, due to their vanishing coupling to the gauge bosons. 

The leading LHC bounds on neutral scalars come from searches for their
conventional decays into pairs of $\tau$-leptons~\cite{Sirunyan:2018zut}, 
and mainly constrain the low mass and high $\tbeta$ region. The low $\tbeta$
region for high scalar masses, in which the scalar predominantly decays 
into pairs of top quarks, is basically unconstrained. This channel remains
an experimental challenge due to the complicated final state, large 
backgrounds and non-trivial interference patterns~\cite{Carena:2016npr}. 
Note that limits from searches for conventional decays are significantly 
weakened once exotic Higgs decay channels are kinematically allowed. The 
ATLAS~\cite{Aaboud:2018eoy} and CMS~\cite{Khachatryan:2016are} searches for 
the exotic decay mode $A/H \to HZ/AZ$ constrain hierarchical 2HDMs with low scalar masses.

Additional constraints for charged Higgs bosons are derived from experimental
searches at the LHC via the $H^\pm \to \tau\nu$ decay mode. A light
charged scalar with $\mC< m_t$ is mostly excluded by the non-observation of
the decay $t\to H^+ b$, although these limits can be weakened at low $t_\beta$ by 
the existence of exotic decay modes~\cite{Kling:2015uba}.  A heavy charged scalar is only weakly constrained
at very large $\tbeta$~\cite{Aaboud:2018cwk,Aaboud:2018gjj,CMS-PAS-HIG-16-031}.
For a detailed discussion of constraints on the charged Higgs, see~\cite{Akeroyd:2016ymd}.

\end{description}


\subsection{Exotic Higgs decays in Hierarchical 2HDMs}
\label{sec:2hdm-bmp}

We have seen that in a 2HDM with heavy scalar masses close to the
aligment limit, the requirements of unitarity and vacuum stability fix the soft
$\mathcal{Z}_2$ breaking term $\mx = m_H^2 \sbeta \cbeta$ and demand the
mass hierarchy $\mH \leq \mA,\mC$. Additionally, electroweak precision
constraints require the mass of the charged scalar to be close to  that of one of the neutral scalars, $\mC \approx \mH$ or $\mC \approx
\mA$. Hierarchical 2HDMs are therefore restricted to be close to the
following two benchmark scenarios:

\begin{description}

\item[BP-A:] $\mA>\mH=\mC$\\
If the charged Higgs $H^\pm$ is mass-degenerate with the heavy CP-even
Higgs \emph{H}, only the exotic decays of the pseudoscalar \emph{A} are allowed
$\left(A\to H^\pm W^\mp/HZ\right)$. Requiring unitarity
additionally imposes an upper bound on the mass splitting:
$5(\mA^2-\mH^2)<8\pi v^2$.

\item[BP-B:] $\mA=\mC>\mH$\\
If the charged Higgs $H^\pm$ is mass-degenerate with the pseudoscalar
\emph{A}, only the exotic decays into the   CP-even Higgs \emph{H} are
allowed:  $H^\pm \to HW^\pm$ and $A\to HZ$. 
In this case, unitarity imposes an upper bound on the mass splitting: 
$3(\mA^2-\mH^2)<8\pi v^2$.

\end{description}

\begin{figure}[t]
\centering
\includegraphics[width=0.48\textwidth]{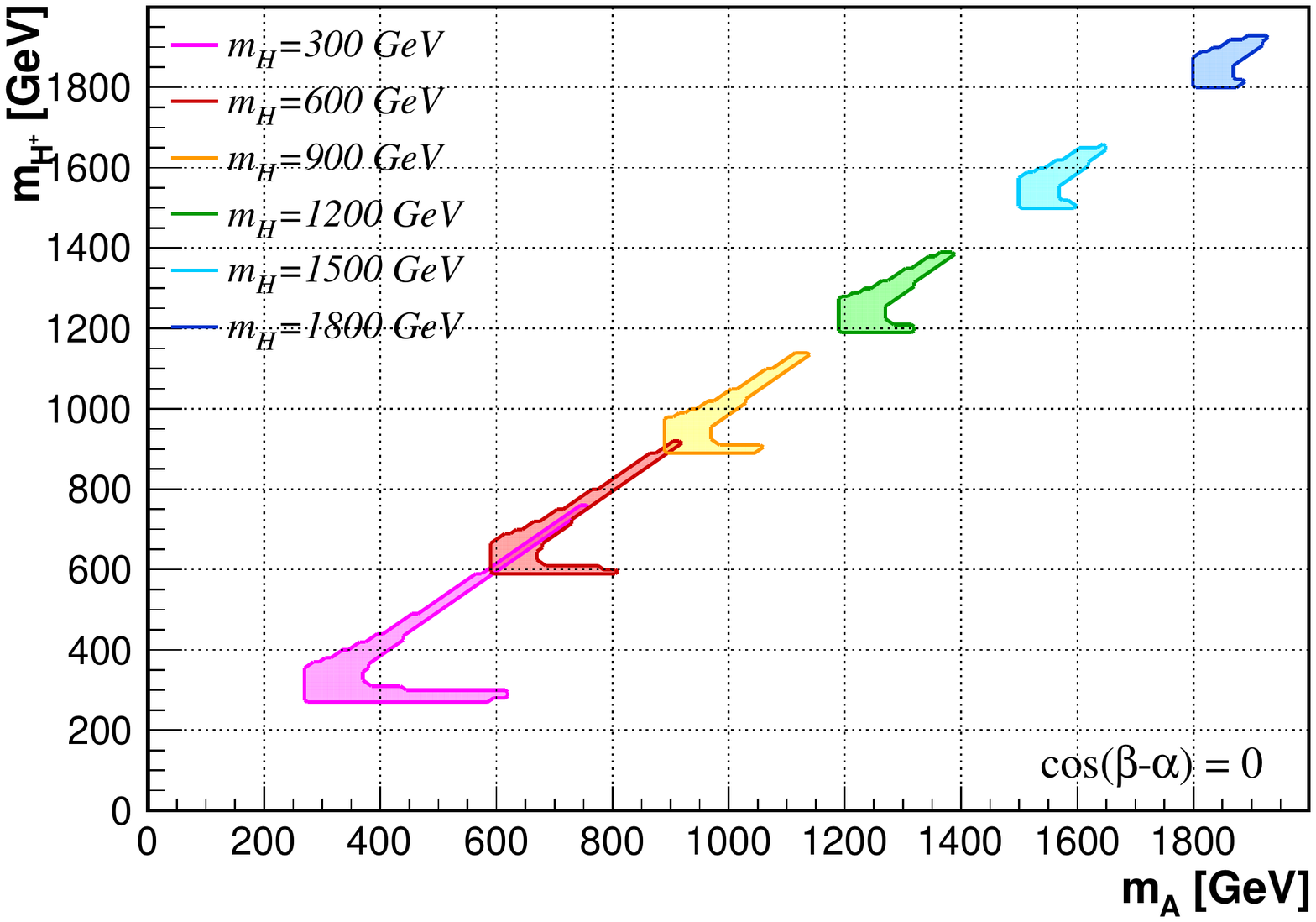}
\includegraphics[width=0.48\textwidth]{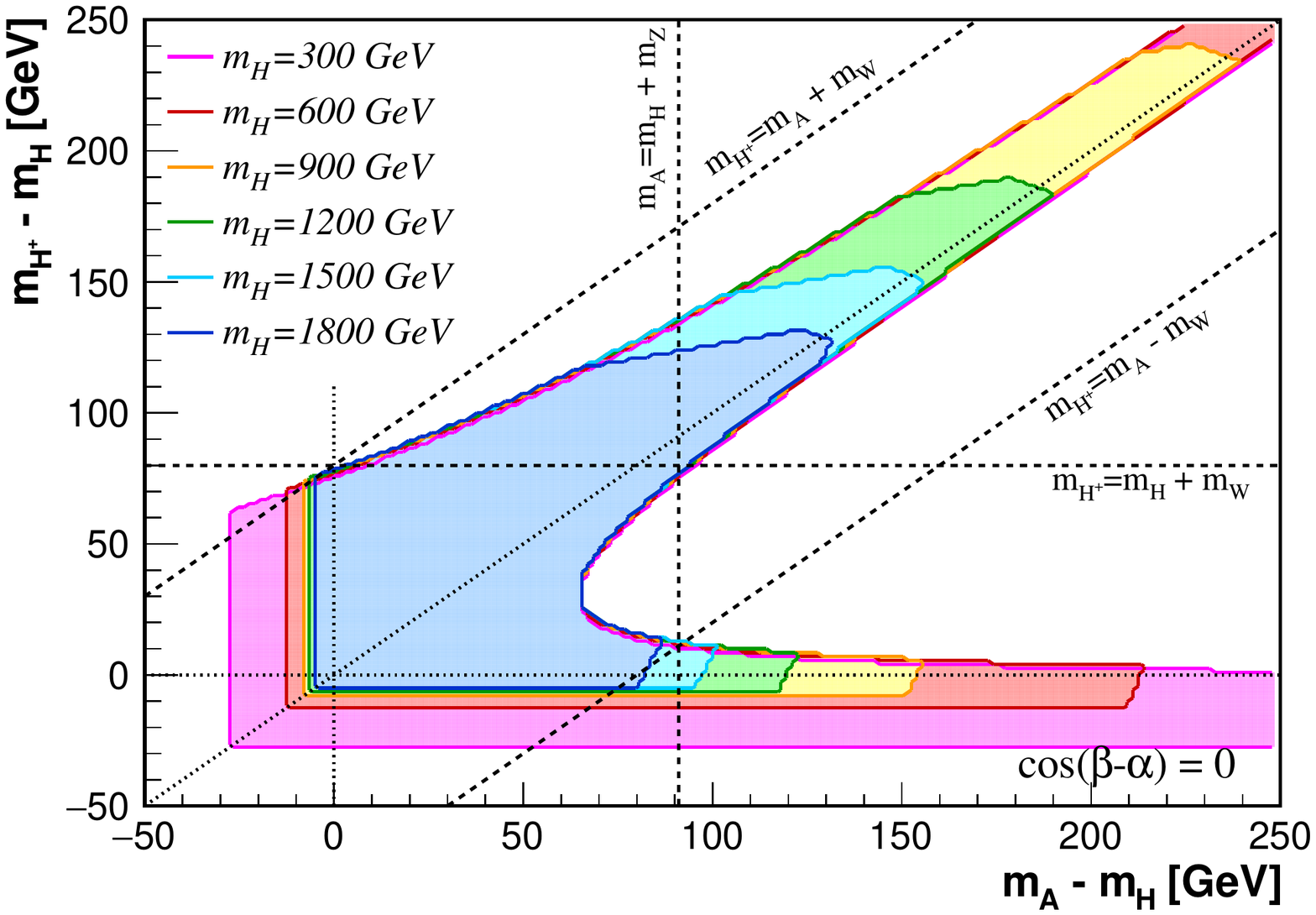}
\caption{
  Allowed regions in parameter space of $m_A$ vs. $m_{H^\pm}$ (left panel) 
  and zoomed-in regions of $m_A-m_H$ vs. $m_{H^\pm}-m_H$ (right panel) 
  considering electroweak constraints, unitarity and vacuum stability for 
  different values of $\mH$. Here we consider the case with $\cba=0$ and 
  $\mx = \mH^2 \sbeta \cbeta$.   
}
\label{fig:constraints}
\end{figure}
While these benchmark scenarios are representative, small
deviations from them are permitted. This is illustrated in
\figref{constraints}, where we show the accessible regions of the
Type-II 2HDM parameter space in the alignment limit when all the theoretical considerations and precision constraints are taken into account.  Note that these
results are independent of the value of $\tbeta$.

While the requirement of vacuum stability imposes a lower bound of $m_H$ 
on $\mA$ and $\mC$, electroweak precision constraints force
the charged scalar to be almost mass degenerate with one of the neutral
scalars. The additional unitarity constraints restrict the mass  
splittings, in particular for large scalar masses, to be small. This
imposes an upper limit on the scalar masses in hierarchical 2HDMs that
permit exotic Higgs decays. The exotic decay channel $A \to HZ$ becomes
kinematically disallowed at $\mA \approx 1.7~\tev$ for \bpa~and $\mA
\approx 2.8~\tev$ for \bpb. Scalar particles in this mass range will be
copiously produced at a future 100 TeV $pp$ collider. Such a machine will
therefore allow us to probe the \textit{entire} hierarchical 2HDM
parameter space, in which the heavy scalar predominantly decays via
exotic modes. For even higher masses, the mass spectrum is forced to be near
degenerate and can be effectively probed by conventional decay channels. Note that
close to the alignment limit, exotic decays of the heavy Higgses into
the light SM-like Higgs $h$, such as $A \to hZ$, $H \to hh$ and $H^\pm
\to h W^\pm$, are suppressed by $\cba$.


\subsection{Production Cross Sections}
\label{sec:2hdm-xs}

In \figref{xs_100}, we show the production cross sections of the
CP-even (left panel), CP-odd (center panel), and charged
(right panel) Higgs bosons at a 100 TeV $pp$ collider as functions of their
masses and $t_\beta$ in the alignment limit. The dominant production
processes for the neutral Higgses are gluon fusion ($gg \to A/H$) and bottom quark
fusion ($bb \to A/H$), shown as solid red and dashed blue lines, respectively. 
The NNLO cross sections for both processes have been calculated using
\textsc{SusHi}~\cite{Harlander:2012pb, Harlander:2002wh,
Harlander:2003ai}. The gluon fusion process will be dominant in the
small $t_\beta$ region, where the production cross section can be greater
than $10^5$ fb for Higgs masses below 600 GeV. In contrast, the
bottom-quark fusion process is dominant in the large $t_\beta$
region. The charged Higgs is predominantly produced via the process 
$gg \to tb H^\pm$, and its production cross section has been adopted from 
Ref.~\cite{Hajer:2015eoa} (which used \textsc{Prospino}~\cite{Beenakker:1996ed, Plehn:2002vy} to calculate it).

Compared to the 14 TeV LHC~\cite{Kling:2016opi}, a 100 TeV $pp$ collider enhances
the production rates of  500 GeV neutral Higgses by roughly 
a factor of 30-50. For charged Higgses with the same mass, the 
rate is enhanced by a factor of 90. For heavier Higgses, the enhancement is even greater.

\begin{figure}[t]
  \centering
  \includegraphics[width=0.32\textwidth]{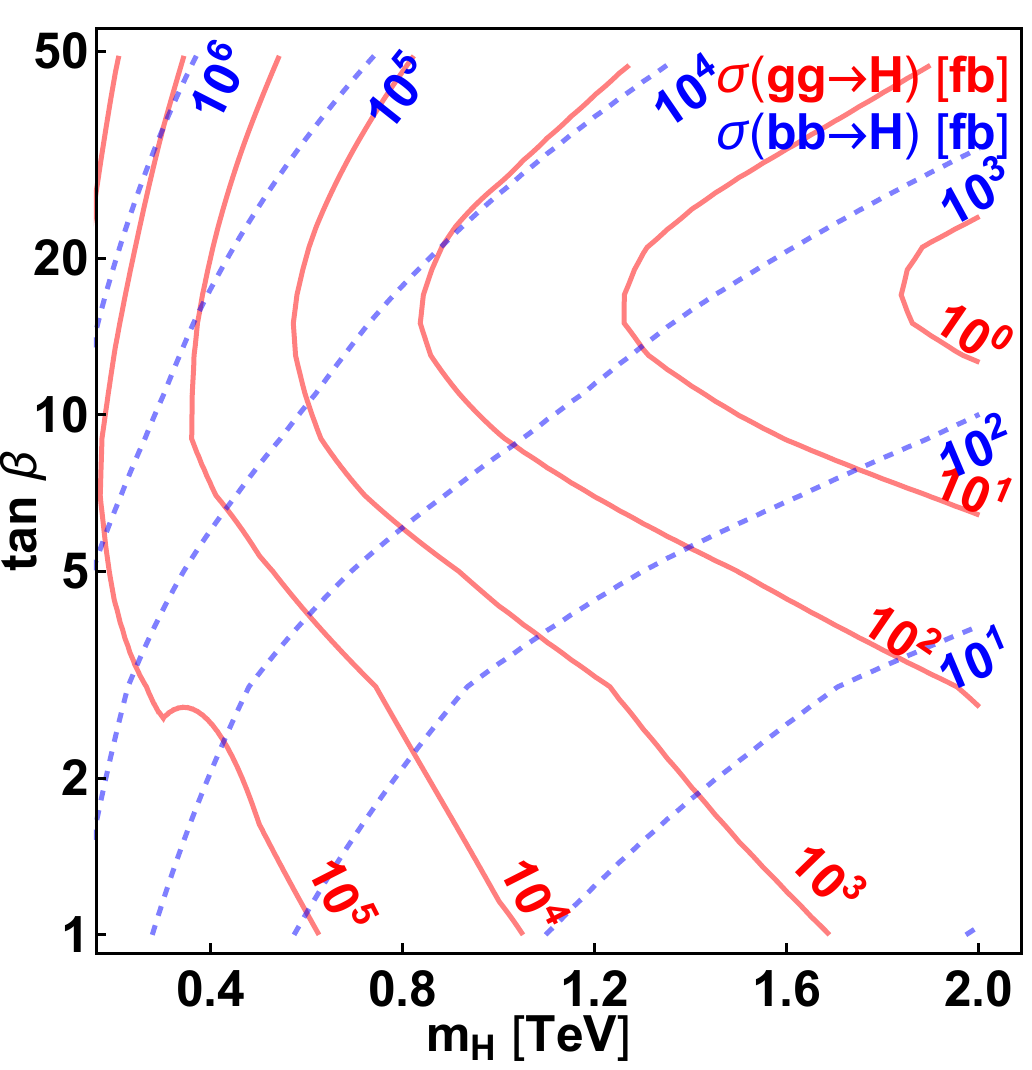}
  \includegraphics[width=0.32\textwidth]{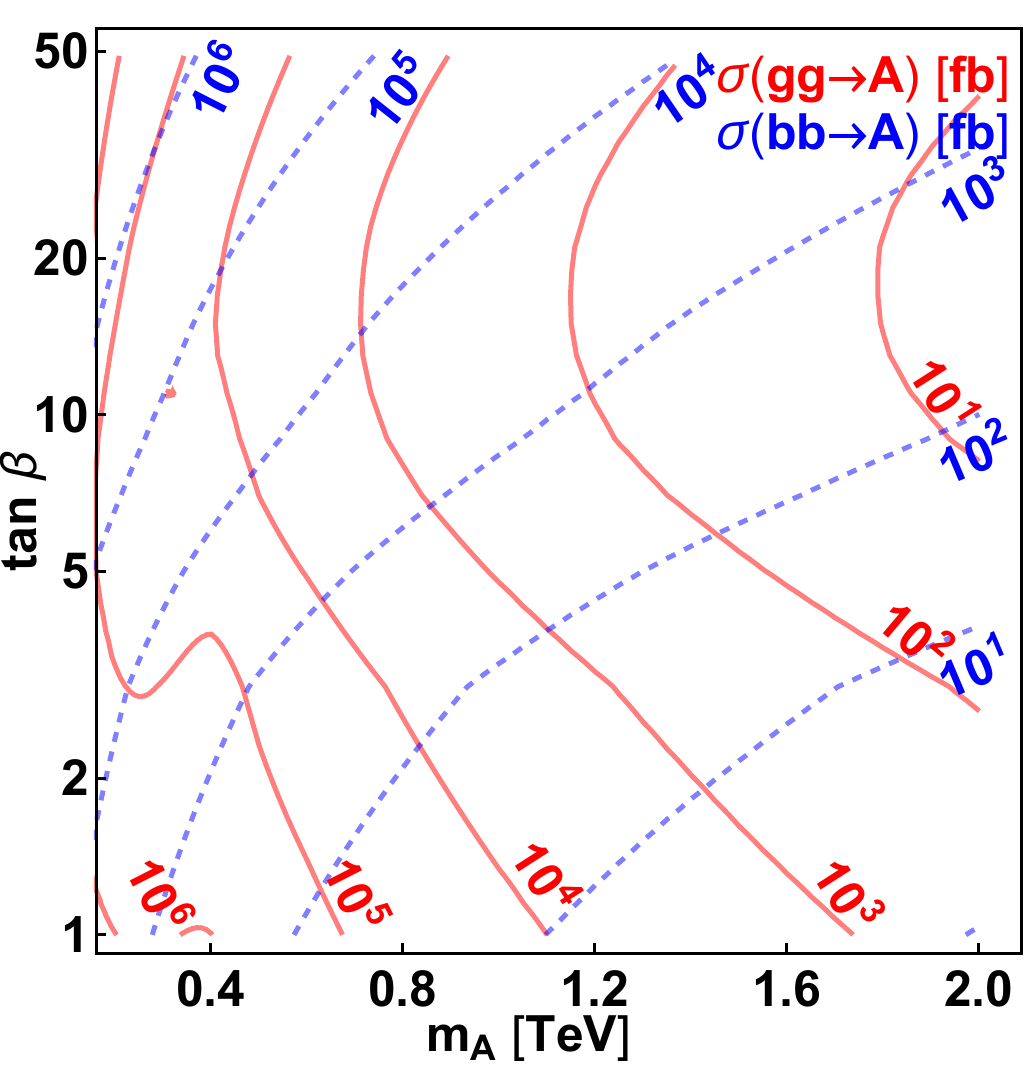}
  \includegraphics[width=0.32\textwidth]{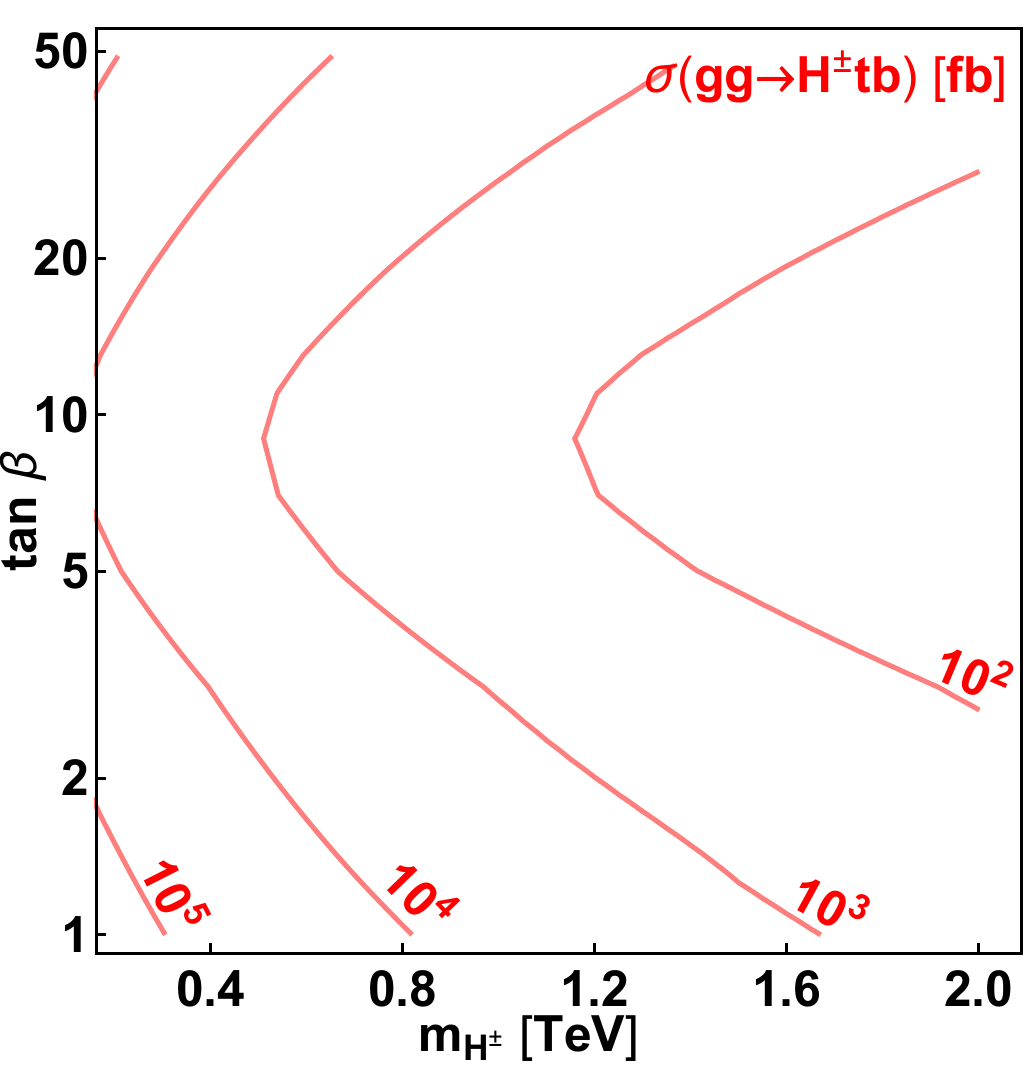}
  \caption{
    Production cross sections for the heavy Higgs bosons \emph{H} (left),
    \emph{A} (center) and $H^\pm$ (right) in a Type-II 2HDM in the alignment limit
    at a 100 TeV $pp$ collider. The red and  blue
    contours correspond to a gluon initial state and a bottom-quark initial 
    state respectively.
  }
  \label{fig:xs_100}
\end{figure}

In \figref{br}, we show the exotic branching fractions of heavy Higgs bosons
as functions of their masses and $t_\beta$ for the two
benchmark scenarios defined in \secref{2hdm-bmp}. The exotic decay
channels have sizable branching fractions ($\gtrsim 20~\%$) over the
entire parameter space and even dominate in the so-called \emph{wedge} region,
corresponding to moderate values of $t_\beta$ ($2 \lesssim t_\beta \lesssim 20$).
This phenomenon reduces the reach of the conventional search channels, but 
also opens up promising avenues for heavy Higgs searches in the form of the 
exotic decay channels. In particular, with the cleanness of the leptonic 
decay modes of the vector bosons, the exotic decays of heavy Higgses provide 
an opportunity to study the wedge region in 2HDMs. 

\begin{figure}[t]
\centering
\includegraphics[width=0.32\textwidth]{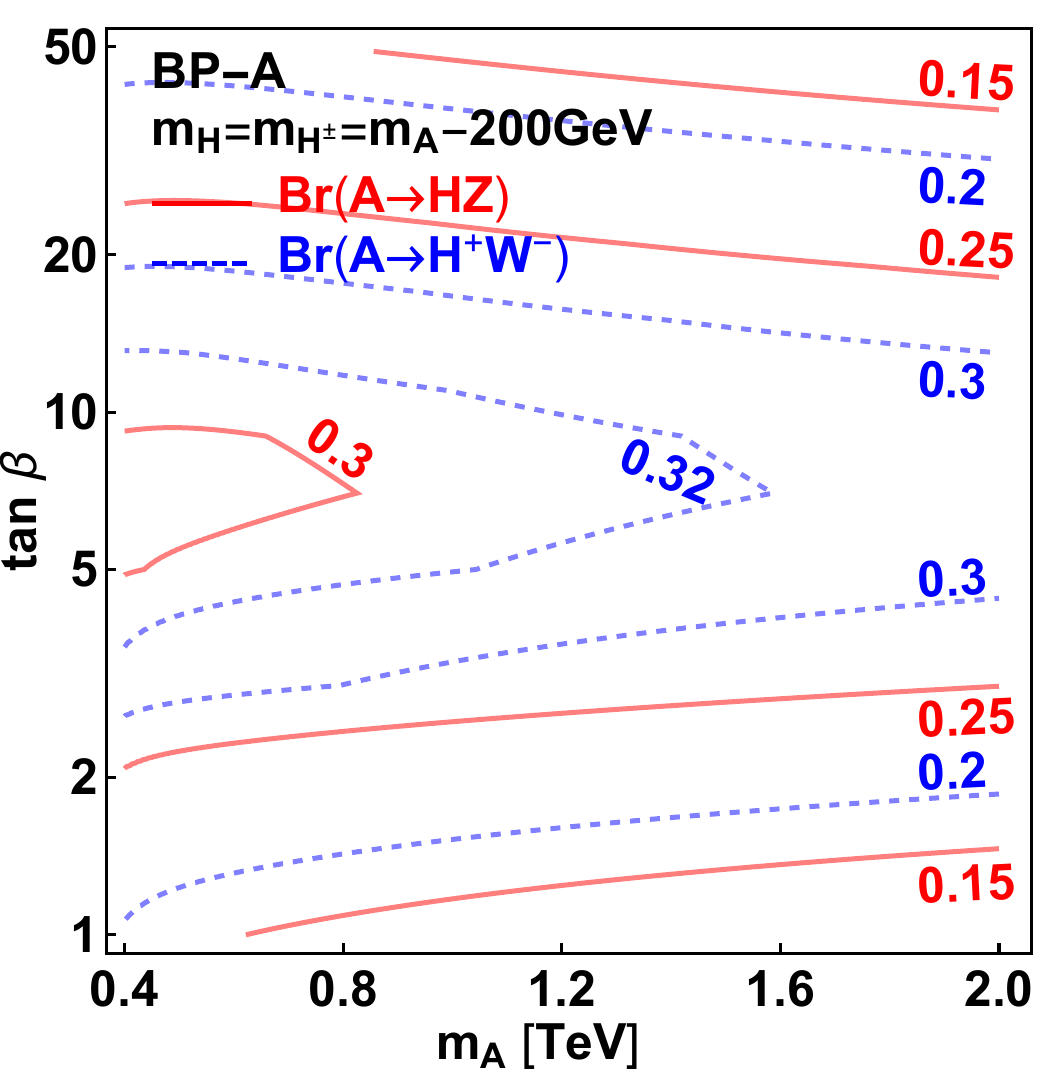}
\includegraphics[width=0.32\textwidth]{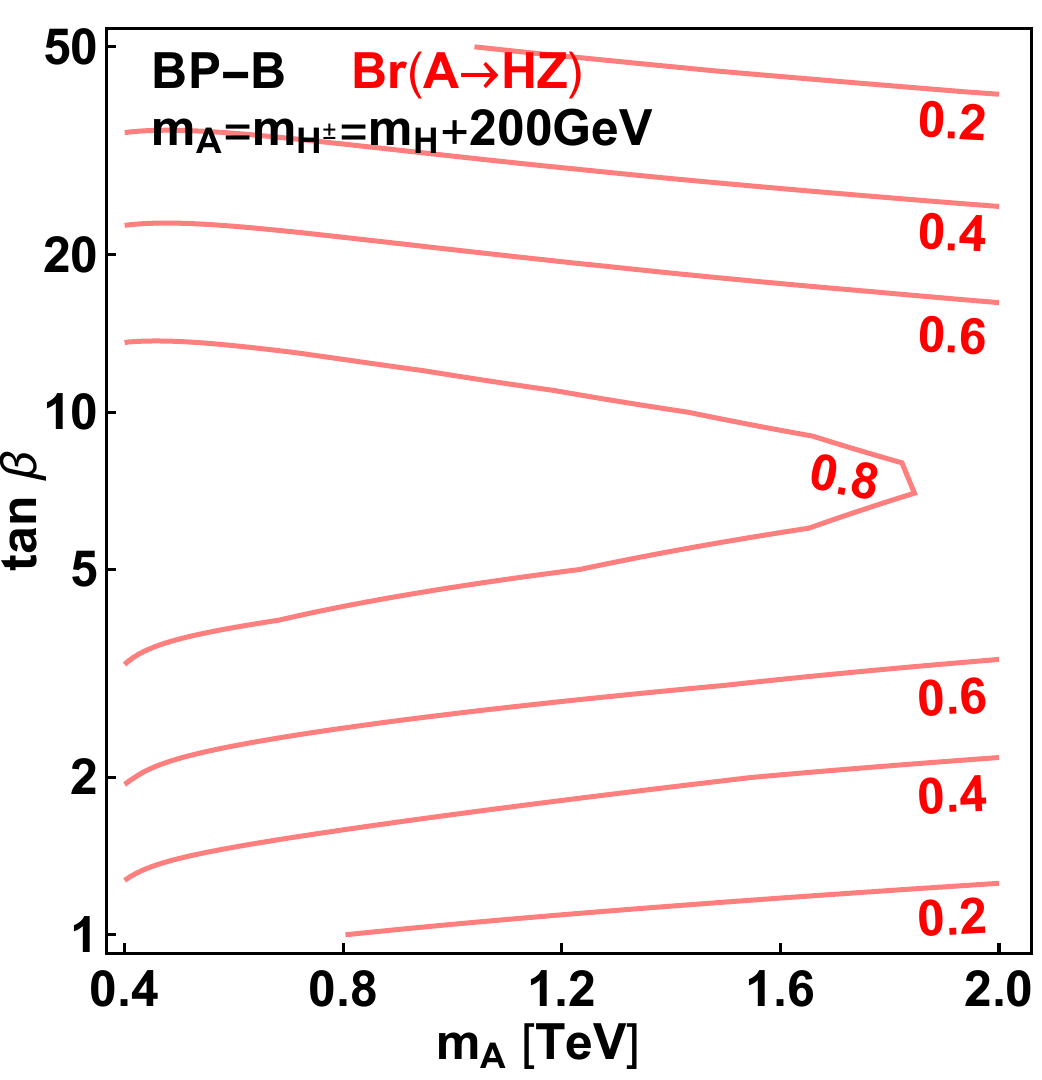}
\includegraphics[width=0.32\textwidth]{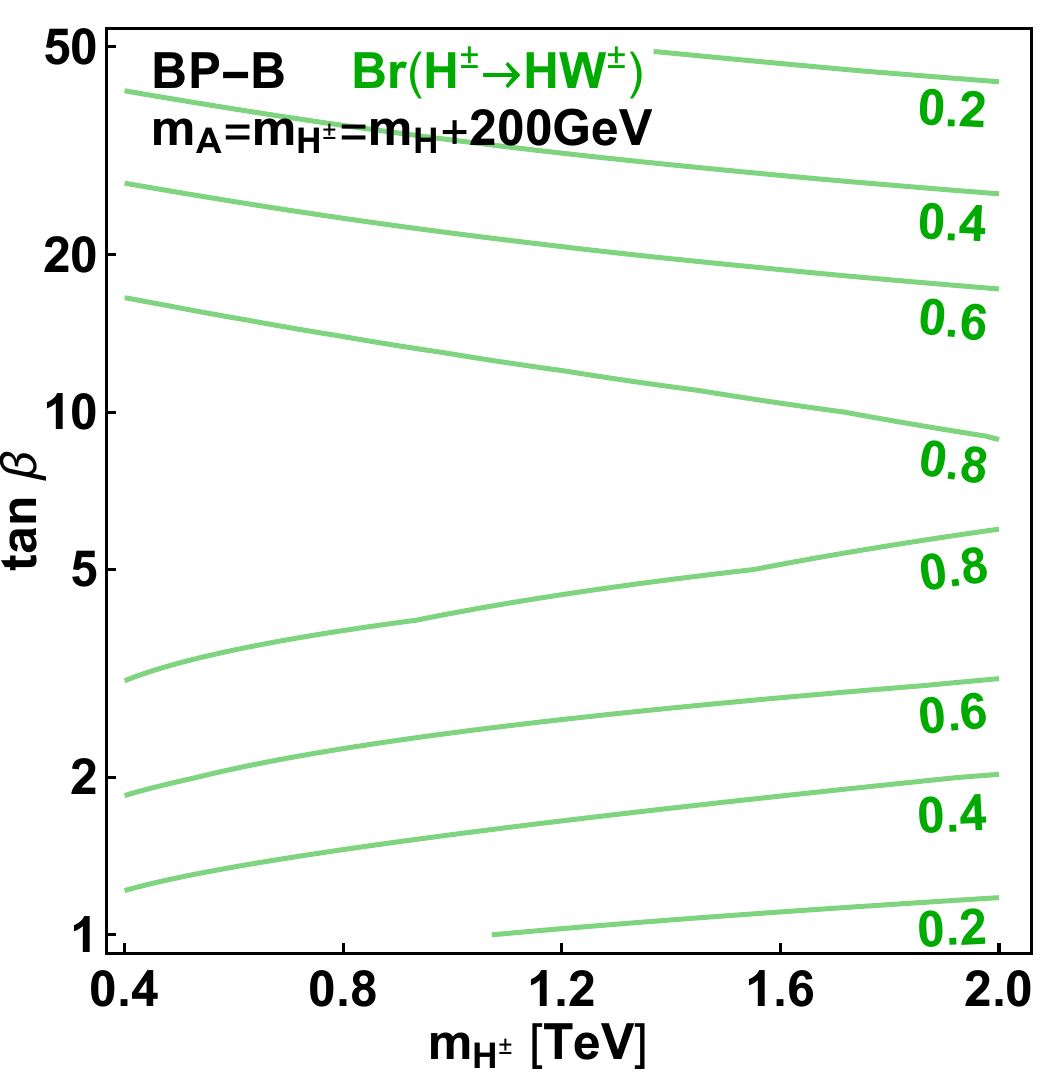}
\caption{Branching fractions for the exotic Higgs decays $A \to HZ$
  (red), $A \to H^+ W^-$ (blue) and $H^\pm \to H W^\pm$ (green). Here we
  consider the benchmark points \bpa~(left) and \bpb~(center 
  and right) with a mass splitting  between the heavy Higgs
  bosons of $\Delta m=200~\gev$.
}
\label{fig:br}
\end{figure}


\section{The Golden Channel: $A \rightarrow HZ$}
\label{sec:HAZ}

\subsection{Signal Processes}
\label{sec:golden_channel_signal_processes}

As discussed in \secref{constraints}, the requirements of unitarity and vacuum
stability constrain the CP-odd state \emph{A} to be heavier than the CP-even
state, thereby opening up the exotic decay mode $A \to HZ$. A further leptonic decay
of the $Z$-boson leads to a experimental signature that is both clean and
covered by the conventional trigger menu of the LHC experiments. This makes
the decay $A \to HZ$ the most promising exotic decay channel, or the \textit{golden channel}.  

Below the top threshold, \emph{H} will predominantly decay to either a
pair of \emph{b}-quarks or a pair of $\tau$ leptons. Although the
branching fraction of the former ($\approx 90\%$) is significantly
higher than that of the latter ($\approx 10\%$), it
suffers  from large SM  backgrounds, making it experimentally
challenging to detect. In contrast,  the latter channel is much cleaner,
making it particularly  interesting at  high luminosities at which
sufficient statistics will be available to make  up for its lower
branching fraction. 

If $m_H$ is above the top threshold, that is, greater than twice
the mass of the top quark, \emph{H} will predominantly decay into top quark pairs
except at large values  of $t_\beta\gtrsim 30$, where the coupling of \emph{H} to
top quarks is suppressed. If $m_H$ is relatively small, leptonic top
decays  will provide the most  sensitive signal. On the other hand, if
it is large,  on the order of a TeV  or greater, the top quarks in the
final state can be highly boosted and top-tagging techniques can be
profitably applied. The latter approach will work  particularly  well
at a future 100 TeV $pp$ collider, at which TeV-scale  heavy
Higgses will be produced in sufficient numbers. In this section we therefore 
consider the three dominant channels  
\be
pp\to A\to HZ\to (bb/\tau\tau/t_h t_h)\ell\ell \ .
\ee
While we focus on the $pp \to A \to HZ$ channel, we note that the
same search can also be performed for the $pp \to H \to AZ$ channel.

\subsection{Analysis}
\label{sec:golden_channel_analysis}


\subsubsection{$bb\ell\ell$-channel}
\label{subsubsec:golden_channel_bbll}

We first consider the $A \to HZ\to bb\ell\ell$ channel, which is the dominant  decay channel for low mass scalars and has been subject to searches at both ATLAS \cite{Aaboud:2018eoy} and CMS \cite{Khachatryan:2016are}.

As discussed in~\cite{Coleppa:2014hxa}, the dominant SM background to this channel is fully-leptonic top pair production ($tt \to bb\ell\ell + \met$), followed by bottom-associated \textit{Z}-boson production ($bbZ \to bb\ell\ell$) for $\ell=e,\mu$.  Decays to $\tau$s are included in the $tt$ background as well.  Additional backgrounds from multi-boson production or mis-tagged jets play a sub-dominant role. The fully-leptonic top pair production background process is simulated with up to one  additional jet and its cross-section normalized to 102 pb and 3714~pb at 14 TeV \cite{Czakon:2013goa} and 100 TeV \cite{Mangano:2016jyj}, respectively. The sub-leading $bbZ \to bb \ell\ell$  background is simulated at leading order taking into  account  a next-to-leading order (NLO) \emph{K}-factor  of $1.45$ \cite{Cordero:2009kv}.  For a transverse momentum threshold of  $p_{b}>15~\gev$, this implies a background  rate of 9.7 pb and 350 pb at 14 TeV and 100 TeV, respectively.

Both the signal and the background process are simulated using \textsc{MadGraph~5} \cite{Alwall:2014hca}, interfaced with \textsc{Pythia} \cite{Sjostrand:2006za,Sjostrand:2014zea} and \textsc{Delphes 3} \cite{deFavereau:2013fsa} for detector simulation. Each signal benchmark is simulated with the correct width and branching fractions as obtained from \textsc{2hdmc} \cite{Eriksson:2009ws}.
We then select events with at least two same-flavor leptons passing the  trigger
requirements $p_{T,\ell_1} > 20~\gev$ and $p_{T,\ell_2}>10~\gev$ and  two $b$-tagged 
jets with $p_{T,b}>25~\gev$\footnote{Stronger selections cuts are applied at a
100 TeV collider for all the search channels (see \appref{method}).}. For these 
events, we construct a set of observables which is then used to train and test a boosted decision tree classifier. For the $bb\ell\ell$ channel, the set of observables includes:
\begin{itemize}
  \tightlist
  \item the transverse momenta of the leading \emph{b}-tagged jet 
  ($p_{T,b_1}$), the  sub-leading \emph{b}-tagged jet ($p_{T,b_2}$), the 
  leading lepton ($p_{T,\ell_1}$) and the sub-leading lepton ($p_{T,\ell_2}$)
  \item  the invariant mass of the leptons ($m_{\ell\ell}$), the jets
  	($m_{bb}$) and the lepton-jet system ($m_{bb\ell\ell}$)
  \item the scalar sum of all the transverse energy ($H_T$) and the missing
    transverse energy ($\met$).
\end{itemize}
Finally, a hypothesis test is performed for each benchmark point to obtain the 
projected statistical significance of the BSM hypothesis versus the SM. We 
assume a 10\% systematic error in the background cross section\footnote{The typical systematic error at the LHC
is between 20\% and 50\%~\cite{Aaboud:2018eoy}.
However, the largest contributions arise from simulation statistics and background 
modeling which could be improved greatly at the future colliders, while theory 
uncertainties are below 10\%.  We adopted a value of 10\% for the systematic uncertainty to
take into account the theory uncertainties.}. More details of our analysis  can be found in \appref{method}. 
 

\subsubsection{$\tau\tau\ell\ell$-channel}
\label{subsubsec:golden_channel_tautaull}

With increasing luminosities, the reach of the $A \to HZ \to bb\ell\ell$ channel
will be limited by systematic uncertainties in estimating the background
rates.  Such limitations do not apply to the $A \to HZ \to \tau\tau \ell \ell$
channel due to its clean final state with significantly smaller background
rates.  Thus, despite having a cross section roughly ten times lower  than that of the $bb\ell\ell$ channel, the sub-leading
$\tau\tau\ell\ell$ channel is  expected to  have a superior reach. 
This channel has been considered by CMS \cite{Khachatryan:2016are} and has
already been found to provide a reach comparable to the $bb\ell\ell$ channel
with the 8 TeV data set. 
In this work we focus on the case in which both $\tau$s decay hadronically, since this allows
for a more precise reconstruction of the Higgs mass than the case in which
one or both $\tau$s decay leptonically, with missing energy arising from neutrinos
in the final state.  Note that the reach can be further enhanced  by combining
the hadronic and leptonic decays, which is beyond the scope of this work. 

The main SM background to the $A \to HZ \to \tau\tau\ell\ell$ signal comes from
boson pair production with the subsequent decay into leptons, 
$(Z/h/\gamma^{*})Z \to \tau\tau\ell\ell$. The corresponding cross sections 
at NLO for the $\tau\tau\ell\ell$ final state are $6.8~\fb$ at 14 TeV \cite{Cascioli:2014yka}
and $67~\fb$ at 100 TeV \cite{Mangano:2016jyj} for 
invariant masses $m_{\tau\tau}>100~\gev$. Note that this includes both resonant
production via \emph{ZZ} and \emph{hZ} dominating at small masses $m_{\tau\tau}$ as well as 
off-shell contributions dominating at large $m_{\tau\tau}$. Sub-dominant backgrounds, 
for example from \emph{ZWW} production, were found to be negligible. 

For this analysis, we select events with  two same-flavor leptons with 
$p_{T,\ell_1} > 20~\gev$ and $p_{T,\ell_2}>10~\gev$ and two $\tau$-tagged 
jets with $p_{T,\tau}>25~\gev$ and consider the following  list of
observables:
\begin{itemize}
  \tightlist
  \item the transverse momenta of leading $\tau$-tagged jet ($p_{T,\tau_1}$),
    the sub-leading $\tau$-tagged jet ($p_{T,\tau_2}$), the leading lepton
    ($p_{T,\ell_1}$) and the sub leading lepton ($p_{T,\ell_2}$)
  \item  the invariant mass of the leptons ($m_{\ell\ell}$), the jets
  $(m_{\tau\tau})$ and the lepton-jet system ($m_{\tau\tau\ell\ell}$)
  \item the scalar sum of all the transverse energy ($H_T$) and the missing
    transverse energy ($\met$).
\end{itemize}


\subsubsection{$tt\ell\ell$-channel}
\label{subsubsec:golden_channel_ttll}

With increasing collision energy, the daughter particle CP-even scalar \emph{H} with
mass above the top threshold can be produced efficiently. In this case, the 
reaches of both the $A\to HZ\to bb\ell\ell$ and the $A \to HZ\to\tau\tau\ell\ell$
channel are limited by statistics due to the suppressed branching fractions, 
especially in the small $t_\beta$ region, while the $A\to HZ\to tt\ell\ell$
channel is expected to improve the reach for \emph{H} above the top quark threshold.
The decay products of \emph{H} can have fairly large $p_T$ for TeV-scale Higgses,
leading to  collimated top decay products. 
Therefore, the standard top reconstruction method for the leptonic decay mode will
lose its efficiency. However, top-tagging techniques~\cite{CMS:2016tvk} developed 
in recent years could retain up to 30\% of hadronic tops while rejecting most of the QCD 
events (see Appendix~\ref{sec:toptagging}). For simplicity, in this work we focus on the case 
in which both tops decay hadronically, which allows for a more 
precise reconstruction of the Higgs mass. Note that mixed hadronic and leptonic
top decays lead to another potentially interesting channel, $A \to HZ\to t_ht_\ell\ell\ell$,
which is beyond the scope of this work.

The dominant SM background to this channel is the process $ttZ\to tt\ell\ell$. The 
corresponding cross section at NLO is $1.91~\pb$ at 100 TeV~\cite{Mangano:2016jyj}.
We select events with two same-flavor leptons passing the trigger requirements 
$p_{T,\ell_1} > 20~\gev$ and $p_{T,\ell_2}>10~\gev $ and two top-tagged jets 
with $p_{T,t}>200~\gev$. The following list of observables is used to train and
test a BDT classifier:
\begin{itemize}
  \tightlist
  \item the transverse momenta of the leading  top-tagged jet
  ($p_{T,t_1}$), the sub-leading  top-tagged jet ($p_{T,t_2}$), the leading lepton
    ($p_{T,\ell_1}$) and the sub-leading lepton ($p_{T,\ell_2}$)
  \item  the invariant mass of the leptons ($m_{\ell\ell}$), the jets
  $(m_{tt})$ and the lepton-jet system ($m_{tt\ell\ell}$)
  \item the scalar sum of all the transverse energy ($H_T$) and the missing
    transverse energy ($\met$).
\end{itemize}


\subsection{Reach}

As discussed in \secref{2hdm-xs}, the production of \emph{A} occurs primarily 
via gluon fusion in the small $\tan\beta$ region and bottom quark fusion in the
large $\tan\beta$ region. We perform a separate analysis for each of these
production modes and combine their significances when presenting the reach.

\begin{figure}[thb]
  \centering
  \begin{tabular}{cc}
  \includegraphics[width=0.45\textwidth]{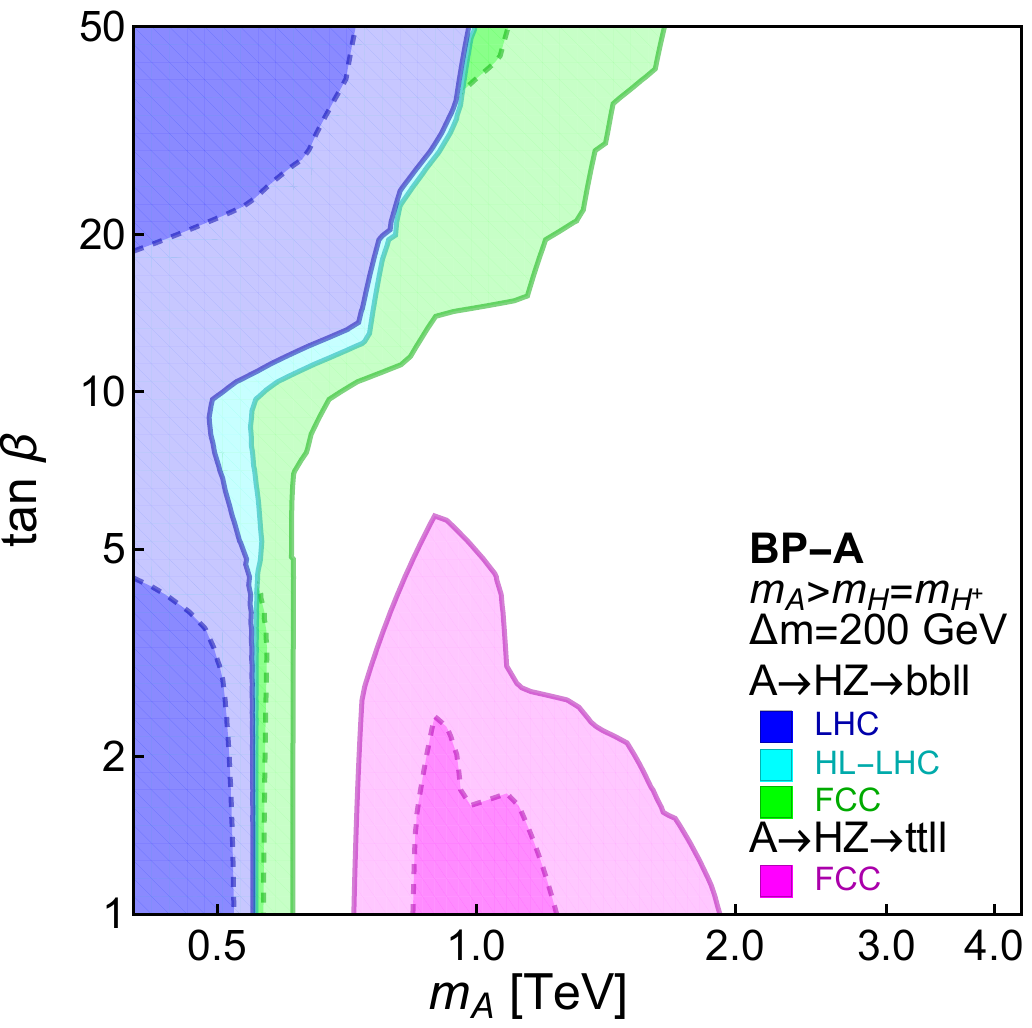}&
  \includegraphics[width=0.45\textwidth]{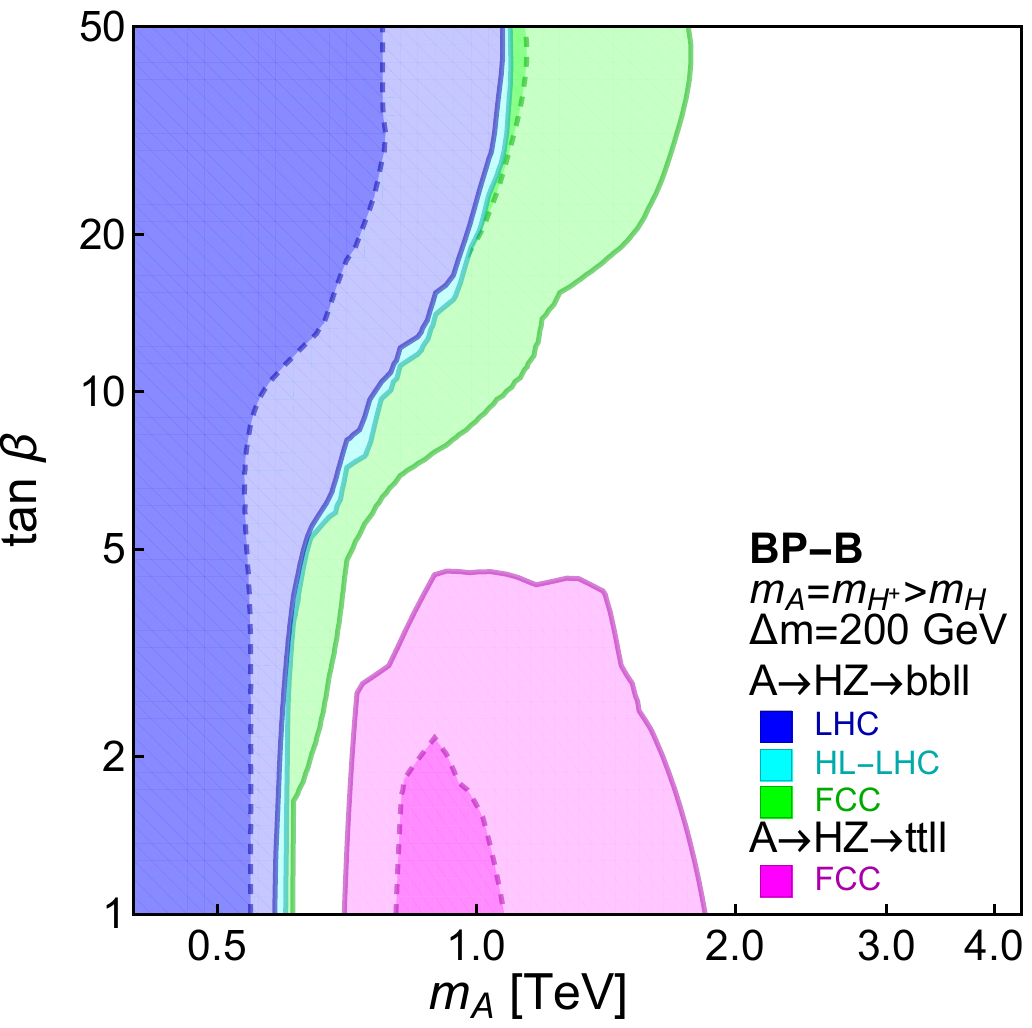}\\
  \includegraphics[width=0.45\textwidth]{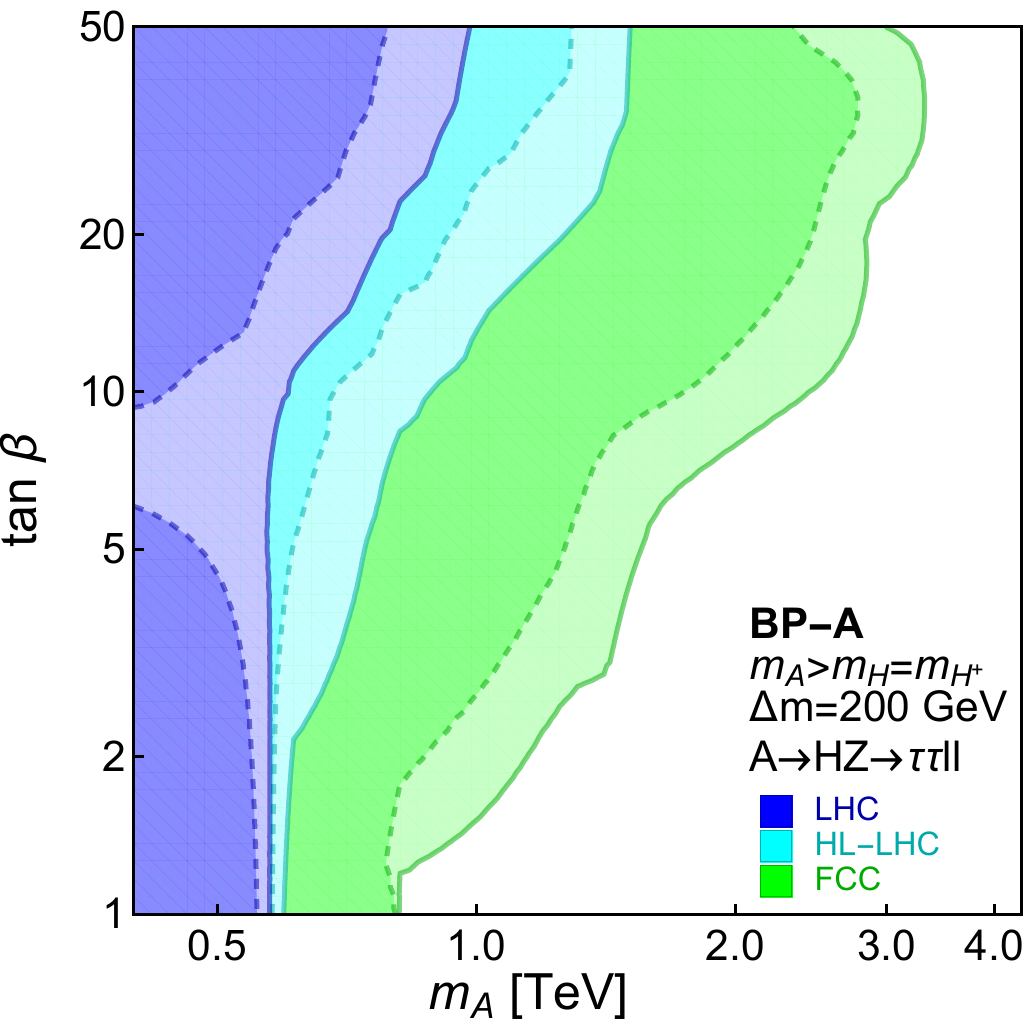}&
  \includegraphics[width=0.45\textwidth]{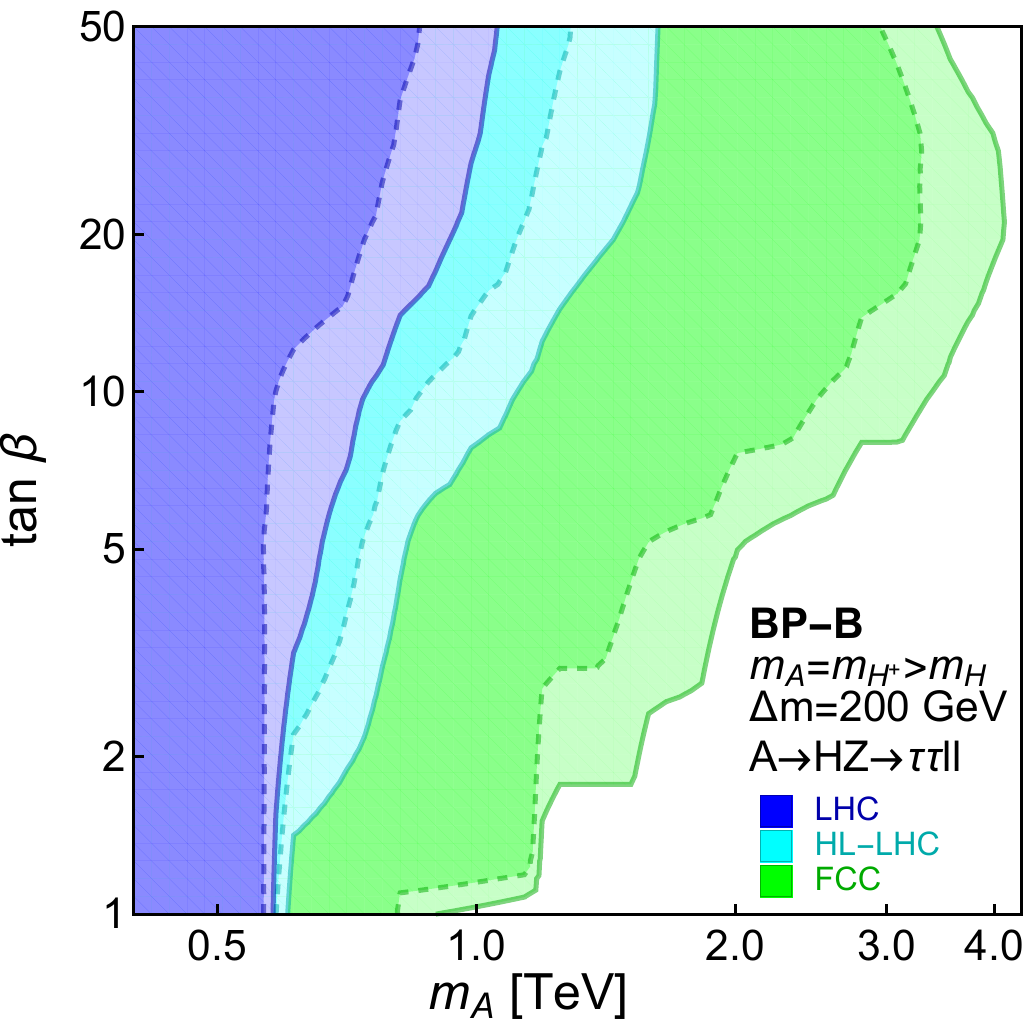}
  \end{tabular}
  \caption{Discovery (dashed) and exclusion (solid) reach for \bpa~(left) 
    and \bpb~(right) at the LHC (green), HL-LHC (cyan) and a 100 TeV  $pp$ 
    collider (blue) in the $\tan\beta$ vs. $m_A$ plane for 
    $m_A-m_H=200~\gev$. 
    We show the reach for the $bb\ell\ell$ and  $t_ht_h\ell\ell$ 
    channels (top), and  $\tau\tau\ell\ell$ channel (bottom).   
     }
  \label{fig:golden_channel_reach}
\end{figure}

In \figref{golden_channel_reach}, we present the discovery (dashed lines) and 
exclusion (solid lines) reach in the $m_A$ vs. $\tan\beta$ plane for \bpa~with 
$m_A=m_{H^\pm} > m_H$ (left panels) and \bpb~$m_A>m_{H^\pm} = m_H$ (right 
panels) at the LHC, HL-LHC and a 100 TeV hadron collider for a fixed 
 mass splitting between the heavy neutral Higgses of $\Delta m =
m_A-m_H=200~\gev$.
The top panels show the reach for the $bb\ell\ell$ and $tt\ell\ell$ final 
states while the bottom panels show the reach for the $\tau_h\tau_h\ell\ell$ 
final state. 

At low values of $\tan\beta$, both the $H \to bb$ and $H \to \tau\tau$ channels are
particularly sensitive at masses below the top threshold, 
$m_A = 2m_t+\Delta m\approx 550~\gev$, while the branching fractions for 
these decays are strongly suppressed at larger masses due to 
the opening up of the $H \to tt$ channel. Increasing the luminosity to 
$3~\iab$ at HL-LHC or a 100 TeV collider does not enhance the reach  
significantly. At large values of $\tan\beta$, the decay $H\to tt$ is 
strongly suppressed and so  the $H \to bb$ and $H \to \tau\tau$ channels  retain sensitivity for large masses. 

The $bb\ell\ell$ channel is limited by systematic uncertainties and hence 
the reach does not increase much  with increasing luminosities 
or center-of-mass energies.
In contrast, the $\tau\tau\ell\ell$ channel has a much 
cleaner signature and therefore is mainly limited by statistical uncertainty
and hence superior in sensitivity to the $bb\ell\ell$ channel. At $\tan\beta=50$ the 
exclusion reach of the $\tau\tau\ell\ell$ channel extends up to 
$\sim 1~\tev$ at the LHC, $\sim 1.5~\tev$ at the HL-LHC and $\sim 3~\tev$ at a 100 TeV $pp$ collider. The maximal discovery regions are around $0.5~\tev$, $1~\tev$ and $2.5~\tev$ for LHC, HL-LHC and 100 TeV $pp$ collider, respectively.

The $H \to t_h t_h$ channel is able to probe scenarios with larger Higgs 
masses in the range $700~\gev \lesssim m_A \lesssim 2~\tev$ for small values 
of $\tan\beta \lesssim 3$.  For smaller masses, the 
sensitivity of this search is limited by the efficiency of the hadronic top-tagging due to smaller typical transverse momenta. At larger values of 
$\tan\beta$,  this search loses sensitivity  due to 
both the smaller Higgs production rates and the smaller Higgs 
branching fraction into top pairs.   

While the heavy pseudoscalar $A$ can decay either into \emph{HZ} or $H^{\pm} W^{\mp}$ in 
\bpa, only the $A\to HZ$ channel is available  in \bpb. Thus, the discovery 
and exclusion reach attainable in \bpb~is greater than in \bpa. 


\section{The Charged Higgs Channel: $A \rightarrow H^\pm W^{\mp} $}
\label{sec:HCW}


\subsection{Signal Processes}

If the mass splitting between the pseudoscalar and charged Higgs is large enough ($\mA > \mC + m_W$), the additional decay channel $A \to H^\pm W^\mp$ opens up. 
This happens in scenarios such as \bpa, where $\mH=\mC<\mA$. In this case
the branching fraction for the exotic decay mode $A\to H^\pm W^\mp$ is 
typically twice as large as that of the $A \to HZ$ decay mode
which can be understood from the Goldstone equivalence theorem. 
The leptonic decay of the \emph{W}-boson provides a clean experimental 
signature and permits the use of a lepton trigger, which makes the decay mode $A \to H^\pm W^\mp$ a promising exotic decay channel to explore.

If the charged Higgs is light  ($\mC \lesssim m_t$),  it 
will dominantly decay into either $\tau\nu$ at high $t_\beta$ or 
$cs$ at low values of $t_\beta$. However, such a  light 
charged Higgs boson is excluded by the non-observation of the top decay 
$t \to H^+ b$~\cite{Aaboud:2018gjj}. If the
charged Higgs is heavier ($\mC > m_t$), the $H^\pm \to tb$ decay mode opens 
up and becomes dominant over the entire phase space. In this case the exotic decay 
channel $A \to H^\pm W \to tbW$ will have the same event topology as top-quark 
pair production, making background suppression the main challenge for this channel. 

If the charged Higgs mass is relatively small ($\mC \sim \text{a few}~100~\gev$), 
the top quark decay products will be both soft as well as  
spread out over the detector area. In this case leptonic top decays are 
expected to provide the most sensitive channel. However, at larger masses  ($\mC \gtrsim 1~\tev$), the top quark from a heavy 
charged Higgs decay will be boosted and top-tagging techniques can be used to 
identify the top quark candidate. In contrast to leptonic top decays, which suffer 
from additional missing energy due to the neutrino in the final state, hadronic 
top decays also allow for a more precise reconstruction of the masses of  the top quark and the charged Higgs. In this study, we therefore 
focus on the following production and decay chain:

\be
  pp \to A \to H^\pm W^\mp \to t_h b\ \ell\nu.
\ee  
 

\subsection{Analysis}
\label{sec:charged_higgs_channel_analysis}

After requiring a hadronic top-tagged jet in the final state,
the leading irreducible background is semi-leptonic top pair production, 
$tt \to t_h b\ell \nu$, where $\ell=e,\mu,\tau$. The corresponding  cross section at a 100 TeV collider is $15.1~\nb$ at NNLO~\cite{Mangano:2016jyj}, which is reduced by a 
factor of roughly 0.2 once we require $p_{T,t}>250~\gev$.
Additional backgrounds arising from the production of a leptonically decaying 
\emph{W}-boson in association with a boosted jet with $p_{T,j}>250~\gev$, which could be 
misidentified as a top quark, were found to be small, 
$\sigma(W^\pm+j\to\ell^\pm\nu+j) = 0.43~\nb$~\cite{Mangano:2016jyj} and are further reduced  upon including the 
mis-tagging rate for QCD jets $\epsilon_j \sim 10^{-3}$ (see 
\appref{toptagging}). Similarly, backgrounds from single top production were
found to be negligible.

We select events containing one lepton with 
$p_{T,\ell_1} > 20~\gev$, at least one top-tagged jet with 
$p_{T, t_1} > 200~\gev$, at least one \emph{b}-tagged jet 
with $p_{T, b} > 50~\gev$ and  a small amount of missing 
transverse energy, $\met > 20~\gev$. The following set of observables is then 
used to train and test a BDT classifier: 

\begin{itemize}
  \tightlist
  \item the transverse momenta of the leading top-tagged jet ($p_{T, t_1}$), 
     the leading \emph{b}-tagged jet ($p_{T,b_1}$) and the leading lepton ($p_{T, \ell_1}$).
  \item the invariant masses of the jets ($m_{tb}$) and the lepton-jet system 
    ($m_{tb\ell\nu}$), and the angular separation of the jets ($\Delta R_{tb}$).
  \item the scalar sum of  the transverse energy ($H_T$) and the missing 
    transverse energy ($\met$).
\end{itemize}

\noindent To reconstruct the mass of  the heavy neutral Higgs ($m_{tb\ell\nu}$), 
we reconstruct the neutrino momentum from $\met$ following the method shown 
in Ref.~\cite{Aad:2015eia}.


\subsection{Reach}

In \figref{A2CW_dm} we present the reach for the exotic decay channel 
$A \to H^\pm W^\pm$ for \bpa{}. Note that this channel is not open in \bpb{}, 
where $\mC = \mA$. We find that the LHC is insensitive to this channel due to 
a low heavy Higgs production rate and  insufficiently 
boosted   decay products. In contrast, a 100 TeV collider will 
be able to produce a sufficient number of heavy Higgses with $\sim \tev$ 
scale masses that can decay  into top quarks with 
the sizable boosts necessary for the use of top-tagging techniques. The 
corresponding exclusion and discovery reaches are  shown as 
solid and dashed lines, respectively. 

\begin{figure}[t]
  \centering
  \includegraphics[width=0.49\textwidth]{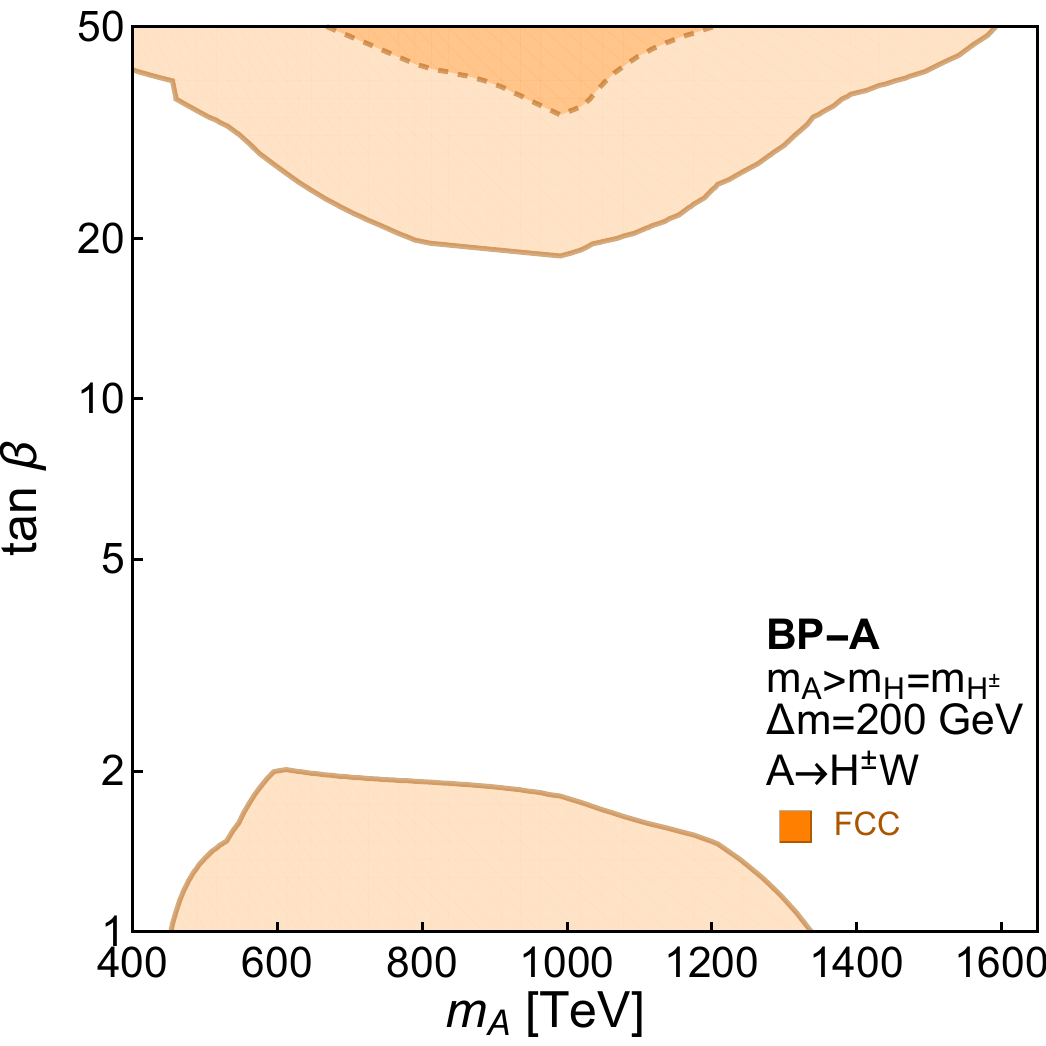}
  \caption{Reach for the exotic decay channel $A\to H^\pm W^\mp\to t_h bl\nu$
    for \bpa~at a 100 TeV $pp$ collider in the $\tan\beta$ vs. $m_A$ plane for 
    $\mA-\mC=200~\gev$. The solid and dashed line correspond to the exclusion 
    and discovery reach, respectively.   
    }
  \label{fig:A2CW_dm}
\end{figure}
 
At small values of $t_\beta$ $(< 2)$ where the pseudoscalar \emph{A} is 
dominantly produced via gluon fusion, the exclusion reach can be up to 
$\mA\simeq 1.3~\tev$. At large $t_\beta$ $(\gtrsim 20)$ the bottom-quark associated production process dominates and this channel can discover a 
CP-odd scalar \emph{A} with mass up to $1.2~\tev$ or exclude CP-odd scalars 
 with masses up to $1.6~\tev$. The 
low reach in the wedge region ($2 \lesssim t_\beta \lesssim 20$), 
results from the small production cross section for both the gluon fusion and 
the bottom quark fusion production of the CP-odd scalar \emph{A}.

Finally, we note that the reach 
of this channel  is dominated by systematic  uncertainties, given the large 
top pair backgrounds. In particular, when estimating the reach we assumed a 
10~\% systematic  uncertainty on the background rate. A better theoretical 
understanding of QCD  processes, especially top-pair production, will be 
extremely important for  accurate background estimation at future 100 TeV 
colliders to reduce the  systematic uncertainties.

\section{Exotic Charged Higgs Decays: $H^\pm \rightarrow HW^\pm$}
\label{sec:CHW}


\subsection{Signal Processes}

While in the previous section we considered exotic decays of neutral Higgses to charged Higgses,
it is also possible for charged Higgses themselves to undergo exotic decays.
As discussed in \secref{2hdm-bmp}, the only viable exotic decay mode for heavy 
charged Higgses in hierarchical 2HDMs in the alignment limit is the decay $H^\pm 
\to H W^\pm$, which appears in \bpb{} when the mass splitting between the charged 
and neutral Higgses is sufficiently large $\left(\mC > \mH + m_W\right)$. As discussed
in \secref{2hdm-xs}, the charged Higgs is mainly produced in association with a 
top and bottom quark $\left(pp \to H^\pm tb\right)$, which leads to a busy final state topology 
$\left(H^\pm tb \to HW^+W^-bb\right)$.

If the daughter Higgs \emph{H} is light $\left(\mH<2m_t\right)$, it will dominantly
decay into pairs of \emph{b}-quarks and $\tau$ leptons with branching fractions
of $\sim 90\%$ and $\sim  10\%$ respectively. Despite its larger branching fraction,
the $H \to bb$ decay channel remains experimentally  challenging, due to the large
hadronic SM backgrounds associated with it\footnote{The authors of \cite{Li:2016umm}
have shown that a jet substructure analysis of the pseudoscalar and 
\emph{W} jets can be used to significantly reduce hadronic backgrounds and provide some 
reach for low values of $m_H$ and $\tan\beta$.}. In contrast, the $H\to\tau
\tau$ decay channel can lead to a same-sign di-lepton signature where one lepton arises from a
leptonic $\tau$-decay and the other from a leptonic \emph{W}-decay. As shown in~\cite{Coleppa:2014cca},
this signature allows for the effective suppression of SM backgrounds - in particular, 
the background from top pair production. 

If the daughter Higgs is heavier $\left(\mH > 2 m_t\right)$, it will dominantly decay into 
pairs of top quarks, leading to a final state equivalent to four top quarks. Searches for 
this channel therefore will be extremely challenging due to the large hadronic SM backgrounds.
However, the authors of~\cite{Patrick:2017ele} have proposed to utilize the possible 
tri-lepton and same-sign di-lepton signatures and have shown that these can be promising 
for larger values of  $m_H$. In this study we consider the following signal 
production and decay chain:

\be
  gg \to H^\pm tb \to H~W^+W^-~bb \to \tau\tau~W^+W^-~bb.
\ee

\noindent with a focus on the same-sign di-lepton final state. 
 

\subsection{Analysis}
\label{sec:exotic_charged_higgs_analysis}

As mentioned above, we consider the case in which one of the \emph{W} bosons and
one of the $\tau$ leptons decay hadronically, and the other \emph{W} boson and 
$\tau$ lepton decay leptonically. The resulting final state permits the same-sign
di-lepton signature $\ell^\pm \ell^\pm + 2b + 2j +\tau_h + \met$, which allows the
suppression of most SM backgrounds. 

The remaining background is dominated by the $tt\tau\tau$ production process, where at least one of the
top quarks decays leptonically (where the definition of leptons includes $\tau$s)~\cite{Coleppa:2014cca}. The 
$\tau$s originate from the decay of a neutral SM boson $\left(Z,h,\gamma^*\right)$. As discussed 
below, the neutral Higgs candidate \emph{H} is reconstructed by combining the momentum 
of the hadronic $\tau$ with the momentum of the softer lepton. A large invariant 
mass of the Higgs candidate in $tt\tau\tau$ background events typically only arises 
when combining a hadronic $\tau$ from boson decay with a lepton from top quark decay,
providing a smooth background spectrum. Using \textsc{MadGraph~5}, we obtain a cross 
section for $t_{h/\ell}t_{\ell} \tau \tau$ production of 886 fb for a 100 TeV 
collider, with the largest individual contribution corresponding to the resonant 
backgrounds \emph{ttZ} and \emph{tth}. For completeness, we also consider the sub-dominant backgrounds, which can provide 
a same-sign di-lepton signature, $ttW\to t_\tau t_\ell\ell\nu$ and 
$ttZ \to t_\tau t_\ell\ell\ell$ with cross sections of 99 fb and 166 fb, respectively. 

Following the analysis strategy outlined in~\cite{Coleppa:2014cca}, we select 
events with two same-sign leptons, one or two \emph{b}-tagged jets, one  $\tau$-tagged 
jet with sign opposite that of the leptons, and at least two untagged jets. We 
loop over all combinations of the untagged jets and choose the combination that
has invariant mass closest to the mass of the \emph{W} boson. We reconstruct the 
leptonically-decaying \emph{W} boson by first reconstructing the neutrino momentum 
using the procedure in \cite{Aad:2015eia} and then combining it with the momentum of 
the hardest lepton. We then combine the momentum of the $\tau$-tagged jet with the 
momentum of the softer lepton to approximate the momentum of the neutral Higgs 
boson \emph{H}. Finally, we combine the \emph{H} candidate with the \emph{W} 
candidate that gives the mass closest to the mass of the charged Higgs. The 
input features for the BDT classifier are the following:

\begin{itemize}
  \tightlist
  \item the transverse momenta of the leading lepton ($p_{T,\ell_1}$), the leading 
  untagged jet ($p_{T,j_1}$), the \emph{b}-tagged jet ($p_{T,b}$), and the 
  $\tau$-tagged jet ($p_{T,\tau_h}$).
  \item the invariant masses of the neutral and charged Higgs candidates 
  ($m_{\tau_h \ell_2}$ and  $m_{\tau_h \ell_2 W}$).
  \item the missing transverse energy ($\met$).
\end{itemize}


\subsection{Reach}

In \figref{C_HW_tataW}, we show the discovery and exclusion reaches (the dashed 
and solid lines respectively) for the exotic decay channel $H^\pm\to HW$ for 
\bpb{}. The reach at the 14 TeV LHC~\cite{Coleppa:2014cca} for 
this channel is limited by the low production cross section of heavy charged 
Higgs bosons, and thus we only show the reach for a 100 TeV $pp$ collider, which 
will be able to produce charged Higgses with TeV-scale masses in large numbers. 

Below the top-quark threshold, $\mA < 2m_t + \Delta m \approx 550~\gev$, the 
$H\to\tau\tau$ channel can probe the entire range of $\tan\beta$. Above this threshold,
the $H\to tt$ decay channel opens up, eliminating the reach at lower values of 
$\tan\beta$. In the interesting wedge region, around $t_\beta=10$, this channel
can discover scenarios with charged Higgs masses up to $1.7~\tev$ and exclude 
charged Higgses with masses up to $2.5~\tev$.

\begin{figure}[t]
 \centering
 \includegraphics[width=0.49\textwidth]{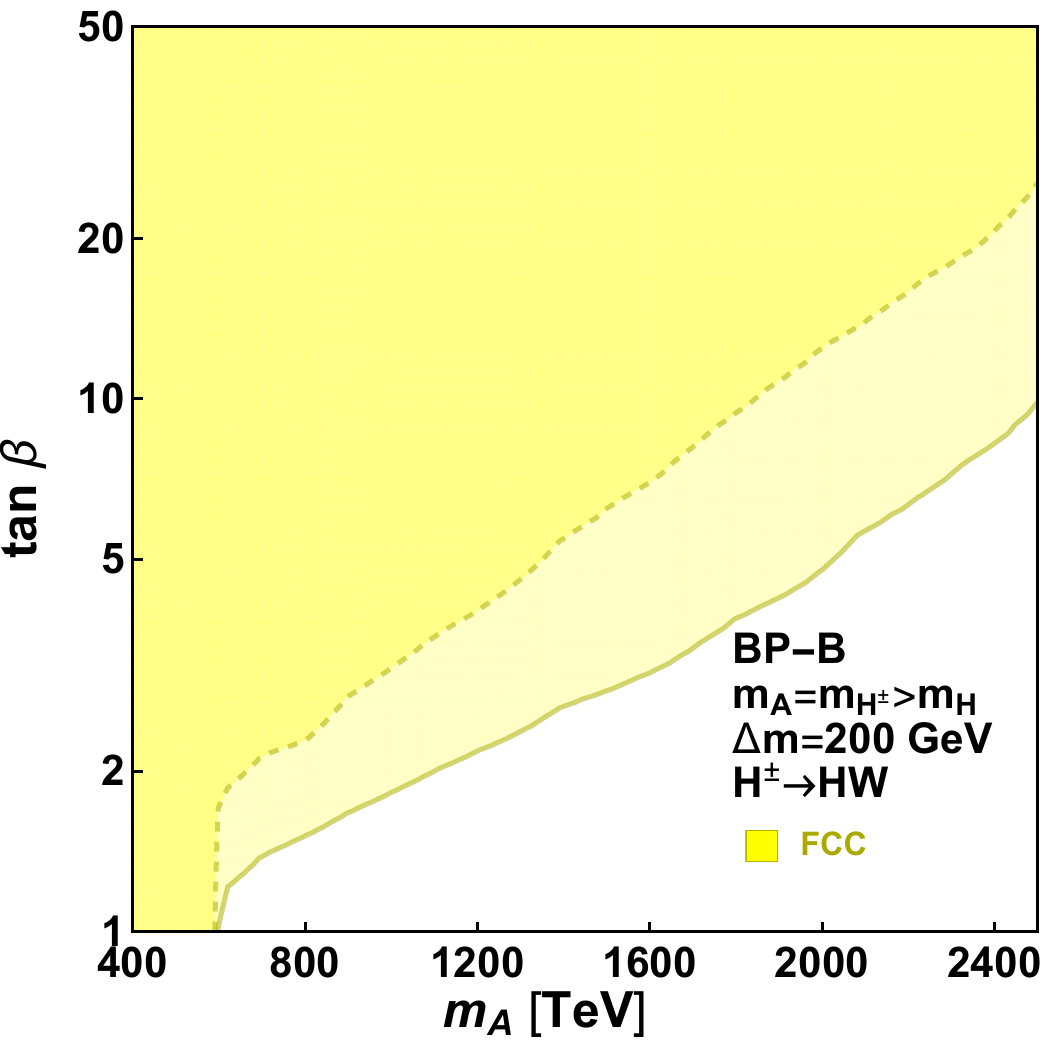}
 \caption{Reach for the exotic decay channel $H^\pm \to H W\to \tau\tau~bb~WW$
    for \bpb~at a 100 TeV $pp$ collider in the $\tan\beta$ vs. $m_A$ plane for 
    $\mA-\mH=200~\gev$. The solid and dashed line correspond to the exclusion 
    and discovery reach, respectively. 
 }
\label{fig:C_HW_tataW}
\end{figure}


\section{Reach in Benchmark Planes}
\label{sec:reach}
 
\begin{figure}[htb]
\centering
\includegraphics[width=0.48\textwidth]{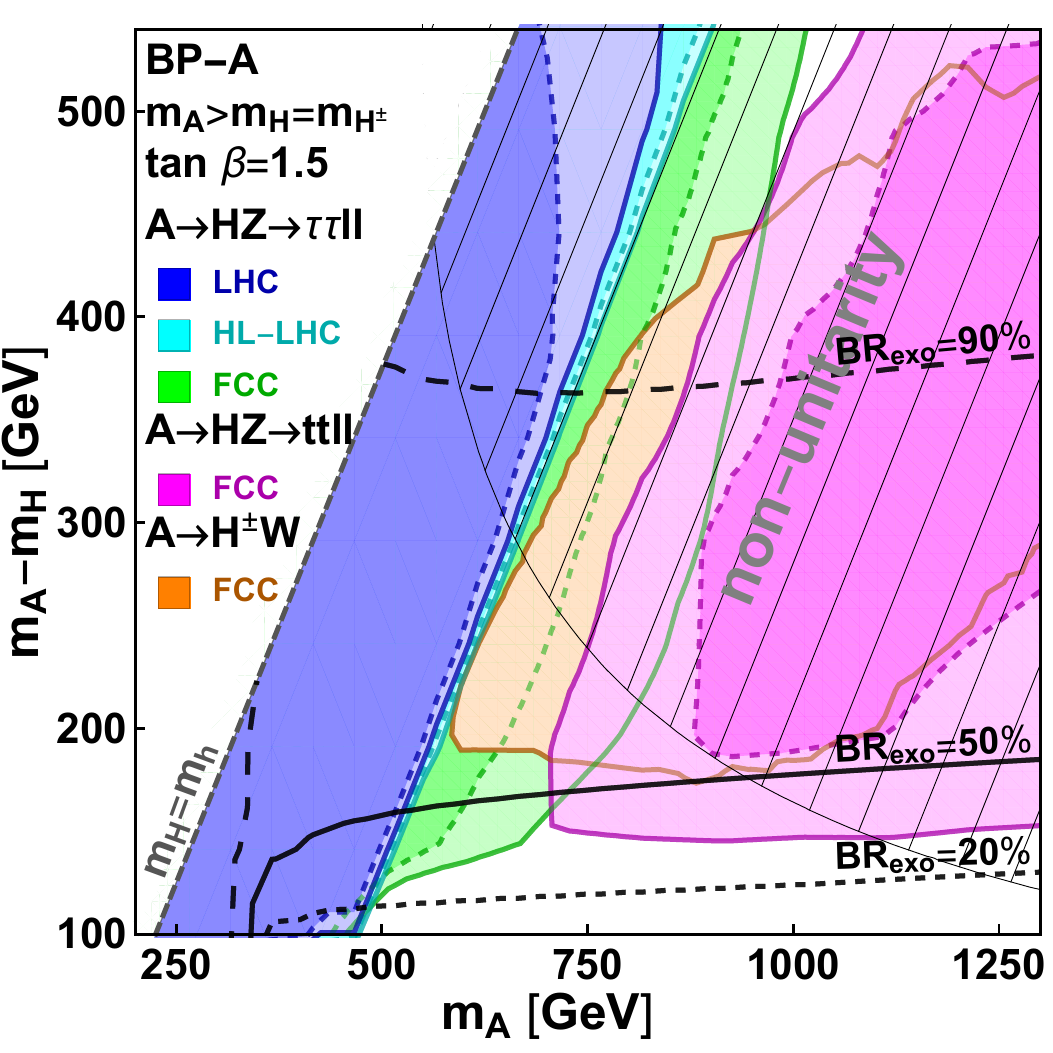}
\includegraphics[width=0.48\textwidth]{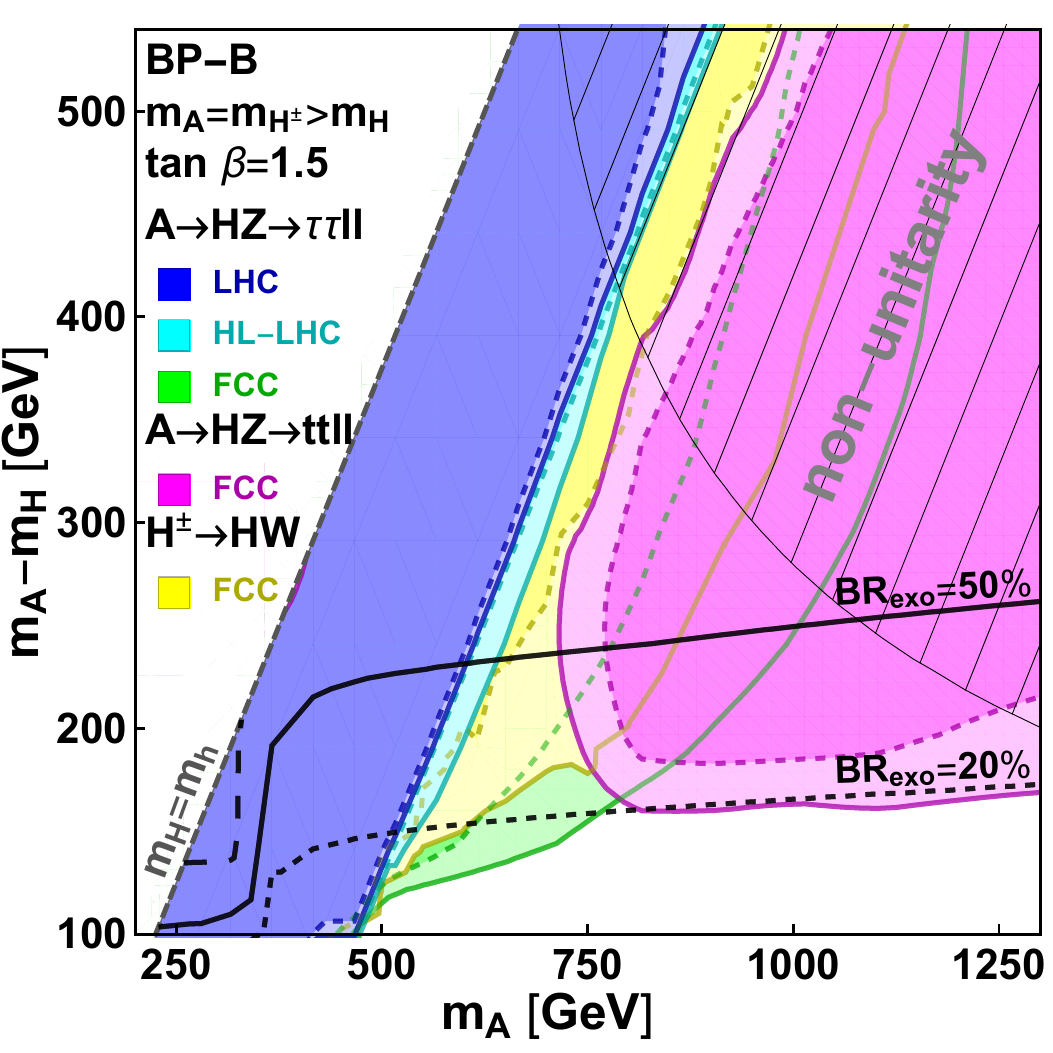}
\caption{Reach for the exotic Higgs decay channels at the LHC, HL-LHC and 100 $pp$ 
TeV collider for \bpa{} (left) with the mass hierarchy $\mH=\mC<\mA$ and \bpb{} 
(right) with the mass hierarchy $\mH<\mC=\mA$. The results are presented in the 
$m_A$ vs. $m_A-m_H$ plane for a fixed value of $\tbeta=1.5$. We show the 
projected sensitivity of the $A \to \tau\tau\ell\ell$ channel (blue/cyan/green)
as discussed in Sec.~\ref{subsubsec:golden_channel_tautaull}, the 
$A \to tt\ell\ell$ channel (magenta) as discussed in 
Sec.~\ref{subsubsec:golden_channel_ttll}, the $A \to H^\pm W^\mp$ (orange) as discussed in Sec.~\ref{sec:charged_higgs_channel_analysis} and the $H^\pm \to HW^\pm$ channel 
(yellow) as discussed in Sec.~\ref{sec:exotic_charged_higgs_analysis}. The 
exclusion and discovery reaches for each channel are shown as solid and dashed 
lines respectively. The hatched regions are excluded by unitarity constraints and 
the thick black lines indicate the branching fraction for exotic Higgs decays 
of the heavy pseudoscalar \emph{A}. 
} 
\label{fig:BPA-B-tb1.5}
\end{figure}

In \figref{BPA-B-tb1.5}, we present the exclusion and discovery reaches in 
the $\Delta m=m_A-m_H$ versus $m_A$ plane for \bpa~(left panel) and \bpb~(right panel) with $\tan\beta=1.5$. As discussed in \secref{2hdm-bmp}, these 
two benchmark scenarios, corresponding to the mass  
hierarchies $\mH=\mC<\mA$ and $\mH<\mC=\mA$ respectively, have been 
found to be representative of hierarchical 2HDMs. In particular, they are 
permitted by theoretical considerations of unitarity and vacuum stability as 
well as electroweak precision measurements. For the purpose of illustration, we consider 
$\tan\beta=1.5$. This choice is representative of the interesting low $\tan\beta$ 
region, which will be particularly hard to constrain using the conventional 
searches such as $A/H \to \tau\tau$ and $H^\pm \to\tau \nu$ which are 
expected to provide the best sensitivity at higher values 
of $\tan\beta$. 

For the $A\to HZ \to \tau\tau \ell\ell$ channel, the  blue, cyan, and  green 
regions show the reaches at the LHC, HL-LHC, and a future 100 TeV collider, 
respectively. For the $A\to HZ \to tt \ell\ell$ channel, as well as the 
channels involving charged Higgs bosons, $A\to H^\pm W$ and $H^\pm\to AW$, 
the reaches at a 100 TeV collider are shown in  magenta, orange and yellow, 
respectively. For each of the six colors, we distinguish between discovery 
and exclusion regions using differing line styles and opacities for the 
contours and the shading of  the regions they enclose. Regions that are more 
opaque and bounded by dashed contours correspond to discovery, and regions 
that are more transparent and bounded by solid contours correspond to 
exclusion (the discovery regions are always subsets of the exclusion regions).
\medskip 

The  highest sensitivity at low values of 
$\mH$ is provided by the $A \to HZ \to \tau\tau\ell\ell$ channel. At both the 
LHC (blue) and HL-LHC (cyan), the reach extends up to $m_H= 2m_t$, resulting 
in almost straight lines for the sensitivity contours. This can be understood 
from the fact that the $H\to tt$ channel quickly becomes dominant once 
it is kinematically accessible in the low $t_\beta$ regions, with a branching 
fraction close to 100\%. Therefore, in this channel, the HL-LHC will not be 
able to improve the expected reach for hierarchical 2HDMs compared to the LHC.
In contrast, a future 100 TeV collider (green) will be able to provide a 
sufficient event rate for the $A \to HZ \to \tau\tau\ell\ell$ channel to 
significantly extend the reach towards higher masses $m_H>2 m_t$, despite the suppressed branching fraction for $H\to\tau\tau$.
Comparing both benchmark planes, the reach for \bpa{} is slightly 
reduced compared to \bpb{} due to the suppressed branching fraction for the 
$A \to HZ$ in the presence of the additional decay channel $A \to H^\pm W$. 
The $A \to HZ \to bb\ell\ell$ channel is limited by systematic errors, resulting in a significantly weakened sensitivity, and is therefore 
not shown in \figref{BPA-B-tb1.5}.
Scenarios with larger Higgs masses $\mH$ can be probed with the decay channel 
$A \to HZ \to tt\ell\ell$. We focus on the case of hadronically decaying top 
quarks, which can be identified using top tagging techniques, and present the 
reach at a 100 TeV hadron collider (magenta). The sensitivity is weakened 
in regions with lower Higgs masses $\mH\lesssim 600~\gev$ in which the
top quarks will no longer have sufficient transverse momentum $\left(p_{T,t} \sim (m_H-2m_t)/2\right)$ 
to exceed the top tagging threshold $\left(p_{T,t}>200~\gev\right)$. As before, the reach 
in \bpa{} is reduced relative to \bpb{} due to the lower branching fraction for 
the decay $A \to HZ$.    
\medskip 

In addition to the neutral Higgs channel $A \to HZ$, hierarchical 2HDMs can 
also be probed via exotic Higgs decays involving charged Higgs bosons. \bpa~ 
permits the additional exotic Higgs decay channel $A \to H^\pm W$. Above the 
top threshold, the charged Higgs decays predominantly into $H^\pm \to tb$. 
Again we focus on subsequent hadronic top decays, which permit the use of top 
tagging techniques, and obtain the projected sensitivity at a 100 TeV 
collider (orange). For smaller charged Higgs masses 
$\left(\mC \lesssim 400~\gev\right)$, the sensitivity of this search 
channel is  limited by the efficiency of the hadronic top-tagging due to 
smaller typical transverse momenta $p_{T,t}\sim (\mC-m_t)/2$. Note that the 
slightly larger typical $p_{T,t}$ in $H^\pm \to tb$ decays 
compared to $H \to tt$ decays results in a mildly extended reach towards 
lower masses compared to the $A \to HZ \to tt\ell\ell$ channel.

The exotic decay of a charged Higgs boson $H^\pm \to H W$ is permitted only 
in the mass hierarchy of \bpb. While searches for this channel at the LHC 
suffer from a low charged Higgs production rate, the production cross section 
increases significantly towards higher energies. We obtain the projected 
sensitivity at a 100 TeV hadron collider (yellow) considering the neutral 
Higgs decay $H \to \tau\tau$. Below the $H\rightarrow tt$ threshold, 
this channel provides 5-$\sigma$ discovery at a future 100 TeV collider, 
which is comparable with $A\to HZ\to\tau\tau\ell\ell$ channel.
\medskip

As discussed in \secref{2hdm-bmp}, unitarity disfavors large mass splittings 
$\mA-\mH$ at large Higgs masses $\mA$. This constraint is  
represented by the hatched region in  \figref{BPA-B-tb1.5}. 
In particular, unitarity constrains a larger region of parameter space 
for \bpa{} than for \bpb{}, imposing upper 
bounds on the mass splittings of  $5(\mA^2-\mH^2)<8\pi v^2$ and 
$3(\mA^2-\mH^2)<8\pi v^2$, respectively.

To indicate the importance of exotic Higgs decays relative to the 
conventional Higgs decays, we also show branching fraction for exotic Higgs 
decays of the heavy pseudoscalar \emph{A} as black contours in \figref{BPA-B-tb1.5}. 
The dotted, solid, and dashed  black contours correspond 
to  branching fractions of 20\%, 50\%, and 90\%,
respectively. We can see that a future 100 TeV hadron collider will be able to 
probe the entire region of the Type-II 2HDM parameter space that survives 
current theoretical and experimental constraints with exotic branching 
fraction $\gtrsim 20\%$ using the combination of all viable heavy Higgs 
exotic decay channels. 

\section{Conclusion }
\label{sec:conclusion}
 
While most direct searches for an BSM Higgs sector focus on the conventional 
decays of the corresponding Higgs bosons, additional exotic decays of these 
states can arise if the BSM Higgs sector is hierarchical. These exotic decays 
include the decay of a heavy Higgs to two lighter Higgses, or to a lighter 
Higgs and a SM gauge boson. The presence of those exotic decay channels 
weaken the bounds of conventional searches, but also open up new 
complementary search channels. 

In this paper, we studied the sensitivity  of the LHC, HL-LHC 
and s 100 TeV $pp$  collider to  exotic Higgs decays in 
Type-II 2HDMs. As 
discussed in \secref{2hdm}, theoretical considerations such as
unitarity and  vacuum stability and experimental limits, e.g.  from
electroweak precision measurements, severely constrain the parameter space of 
hierarchical 2HDMs. Besides the fully degenerate case 
$\mH \approx \mA \approx \mC$, there are two benchmark planes that are viable 
under the alignment limit: \bpa~($m_A > m_H=m_{H^\pm}$) with $A\rightarrow 
HZ/H^\pm W^\mp$ and \bpb~($m_{A}=m_{H^\pm}>m_{H}$) with 
$A\rightarrow HZ$, $H^\pm\rightarrow H W^\pm$.

A 100 TeV $pp$ collider provides the opportunity to probe exotic decays of 
heavy Higgses with top quarks in the final state. Top quarks originating from 
the decay of a heavy Higgs are typically boosted, permitting the use of top 
tagging techniques to identify them. This allows us to take advantage of the 
large decay rates of heavy Higgses into top quarks while also 
 getting a handle on   QCD backgrounds. 

To obtain the projected reach of the considered exotic Higgs decay channels, 
we perform a multivariate analysis using  boosted decision tree classifiers 
which are trained to  distinguish between the signal  events
and   the SM background  events. We find that the best 
sensitivity is provided by the exotic decay channel $A \to HZ$ due 
to its clean final state,  and hence we term it the \textit{golden} channel. Regions of parameter space with  low values of $\mH$ $\left(\mH<2m_t\right)$ and 
large values of $\tan\beta$ can efficiently be probed with the final states $
bb\ell\ell$ and $\tau\tau\ell\ell$, where the $\tau\tau\ell\ell$ channel has 
a  better reach compared to $bb\ell\ell$ channel 
due to the significantly lower backgrounds. For moderate mass 
splittings $\left(\mA-\mH=200~\gev\right)$ and large values of $\tan\beta$ 
$\left(>10\right)$, a 100 TeV \emph{pp} collider can discover (at 
$5\sigma$) and exclude (at 95\% C.L.) Higgs masses up to $\mA \approx 3~\tev$ 
and $4~\tev$, respectively. In the  low $\tan\beta$ region above the top-pair 
threshold, the $tt\ell\ell$ channel is complementary to $\tau\tau\ell\ell$, 
extending the reach to about $\mA \approx 1.2~\tev$ ($2~\tev$) for discovery (exclusion).

Hierarchical 2HDMs can further be probed via exotic decay channels involving 
the charged Higgs boson. In the mass hierarchy  corresponding to \bpa{},
the exotic decay channel $A \to H^\pm W$ is kinematically open. Using the 
dominant charged Higgs decay mode $H^\pm \to tb$, a 100 TeV 
collider can exclude Higgs masses up to $\mA \approx 1.6~\tev$ at large 
$\tan\beta$ $\left(\approx 50\right)$ and about $\mA \approx 1.3~\tev$ at 
small $\tan\beta$ $\left(\approx 1\right)$ for a mass splitting of 
$\mA-\mH=200~\gev$. In \bpb{}, exotic decays of the charged Higgs 
$H^\pm \rightarrow H W$ become kinematically permissible. We analyze this decay considering $tbH^\pm$ associated charged 
Higgs production and  the subsequent decay of the neutral Higgs   $H \to \tau\tau$, which permits for a same-sign di-lepton signature. For moderate mass splittings 
$\left(m_A - m_H=200~\gev\right)$ and values of $\tan\beta$ 
$\left(\approx 10\right)$, a 100 TeV $pp$ collider  can discover (exclude) 
Higgs masses up to $\mC \approx 1.7~\tev$ and $2.4~\tev$, respectively. The 
channel $H\rightarrow tt$ could provide additional reach  at low values of $\tan\beta$ 
above the top pair threshold~\cite{Patrick:2017ele}.
 
Combining all  the aforementioned  exotic decay 
channels, we present the  reach in the benchmark planes \bpa{} and \bpb{} for 
$\tan\beta=1.5$ in \figref{BPA-B-tb1.5}. All three channels complement each 
other nicely: final states with $\tau$s  prove to be  the most sensitive channels for regions with relatively low values of $\mA$, and, as might be expected, final states with tops are 
useful above the top threshold. We find that these exotic Higgs decay 
channels can probe the entire parameter space in  which the exotic decay 
branching fraction is more than 20\%. Additionally, if  a future 100 TeV 
collider observes the $A \to HZ$ channel, it would imply the existence of additional exotic decay
channels involving the charged Higgs, which will be observable in many parts of the parameter space. 
 
While most of the recent searches for additional Higgs bosons have focused on 
conventional decay channels, searches using exotic decay channels have just 
started \cite{Aaboud:2018eoy, Khachatryan:2016are}.  At a possible high energy future hadron collider, both the exclusion and the discovery reach for 
non-SM Higgses will be greatly enhanced compared to that of the LHC. The 
discovery of a  non-SM heavy Higgs would  serve as 
unambiguous evidence for new physics beyond the SM and could also provide 
valuable insights into mechanism  underlying electroweak 
symmetry breaking.  

\begin{acknowledgments}

We would like to thank  Ahmed Ismail for providing the production cross 
section of the charged Higgs at a 100 TeV $pp$ collider. An allocation of computer time from the 
UA Research Computing High Performance Computing (HPC) and High Throughput 
Computing (HTC) at the University of Arizona is gratefully acknowledged. FK 
is supported by the U.S.  National Science Foundation under the grant 
\texttt{PHY-1620638}. AP, HS, and SS  were 
supported by the Department of Energy under Grant \texttt{DE-FG02-13ER41976/DE-SC0009913}. HL was supported  by  the National Natural Science 
Foundation of China (NNSFC) under grant No. \texttt{11635009} and Natural Science Foundation of 
Shandong Province under grant No. \texttt{ZR2017JL006}. 

\end{acknowledgments}

\appendix
\section{Collider Analysis Methodology}
\label{sec:method}

In this section, we describe the details of the methodology we employ for our 
collider analysis: For each set of considered model parameters, we generate 
Monte Carlo event samples for both signal and background processes, train a BDT 
classifier to distinguish between signal and background events and perform a hypothesis test to 
obtain the expected statistical significance.

The production cross sections for the heavy pseudoscalar \emph{A} are 
calculated using \textsc{SusHi} \cite{Harlander:2012pb, Harlander:2002wh, 
Harlander:2003ai} at NNLO. The charged Higgs productions rates have been 
adopted from \cite{Hajer:2015eoa} and therein were calculated\footnote{We thank Ahmed Ismael for providing  us with the production cross sections for the charged
Higgs.} using \textsc{Prospino}~\cite{Beenakker:1996ed, Plehn:2002vy}.
The decay width and branching fraction for each simulated signal benchmark point is calculated using 
the \textsc{2hdmc} package \cite{Eriksson:2009ws}. 

We simulate parton-level events using \textsc{MadGraph~5} and \textsc{MadEvent}
\cite{Alwall:2011uj, Alwall:2014hca} with a modified 2HDM model,
2HDM-HEFT \cite{Degrande:2014vpa}, created using FeynRules. This is followed by 
showering and hadronization using \textsc{Pythia} \cite{Sjostrand:2006za,Sjostrand:2014zea},
and fast detector simulation using \textsc{Delphes 3}~\cite{deFavereau:2013fsa}.
For the 14 TeV LHC and HL-LHC scenarios, we used the default \textsc{Delphes} 
detector cards in \textsc{MadGraph}. For the 100 TeV scenario, we used the 
\textsc{Delphes} detector card devised by the FCC-hh working group~\cite{FCC-Delphes-Card}.
In particular, we adopt the following basic selection cuts for detector reconstruction from the  \textsc{Delphes} cards listed above:
\be
  \textbf{LHC/HL-LHC:} & 
  &p_{T, \ell}>&~10~\gev,
  &p_{T,j/b/\tau}>&~20~\gev,
  &\Delta R>&~0.5, 
  \\ &
  &|\eta_{\ell}|<&~2.5,
  &|\eta_{j}|<&~5.0,
  &|\eta_{b/\tau}|<&~2.5  \\
  \textbf{100 TeV:} & 
  &p_{T,\ell}>&~20~\gev,
  &p_{T,j/b/\tau}>&~50~\gev, 
  &\Delta R>&~0.3, 
  \\ &
  &|\eta_{\ell}|<&~6.0,
  &|\eta_{j}|<&~6.0,
  &|\eta_{b/\tau}|<&~6.0  \  
\ee
where $\Delta R$ is the angular distance between any two objects.

The reconstructed-level events from \textsc{Delphes} are filtered through a 
series of trigger and identification cuts (described in sections 
\ref{sec:golden_channel_analysis}, \ref{sec:charged_higgs_channel_analysis}, and 
\ref{sec:exotic_charged_higgs_analysis}), after which a set of  features were 
collected for each simulated collision event to serve as inputs to gradient 
boosted decision tree (BDT) classifiers \cite{Yang:2005nz} implemented in 
\textsc{TMVA} \cite{Hocker:2007ht}. The set of input features included both 
low-level features such as the transverse momenta of individual particles, and 
physically-motivated high-level features such as the invariant masses of 
combinations of particle momenta. The events were then divided into training 
and test sets, and we trained our classifiers on the training sets with the 
following hyperparameters:

\begin{itemize}
  \tightlist
  \item The number of trees was set to 1000.
  \item The maximum depth of each tree was set to 3.
  \item Bagging was employed, with the bagged sample fraction set to 0.6.
  \item The Gini index was used as the separation criterion for node splitting.
\end{itemize}

\noindent The classifiers were then used to compute the BDT response value 
for signal and background events in the test set. We then scanned across a 
range of response values to determine the optimal cutoff with corresponding 
values of the total number of leftover signal (\emph{s}) and background 
(\emph{b}) events that resulted in the greatest discovery and exclusion significance.
The values of \emph{s} and \emph{b} were obtained by multiplying their 
respective cross-sections by the integrated luminosity, which was taken to be 
$300~\ifb$ for the LHC, and $3000~\ifb$ for the HL-LHC and the 100 TeV collider.

Generating a large enough number of Monte Carlo events to estimate the 
backgrounds at a 100 TeV collider was a technically challenging task. For 
certain points in  parameter space, a series of cuts could reduce the number 
of expected background events to zero. However, in such cases, we 
artificially set a minimum three background events, i.e. $b = 3$, to ensure that our significance estimates are not overly optimistic.

To estimate the median expected discovery and exclusion significances, 
$Z_\text{disc}$ and $Z_\text{excl}$, we follow 
\cite{Kumar:2015tna,Cowan:2010af, Cowan:2010js} and use the following  expressions:

\be
  Z_\text{disc} &=
  \sqrt{2\left[(s+b)\ln\left(\frac{(s+b)(1+\epsilon^2 b)}{b+\epsilon^2 b(s+b)}\right) -
  \frac{1}{\epsilon^2 }\ln\left(1+\epsilon^2\frac{s}{1+\epsilon^2 b}\right)\right]} \\
  Z_\text{excl} &=\sqrt{2\left[s-b\ln\left(\frac{b+s+x}{2b}\right) 
  - \frac{1}{\epsilon^2 }\ln\left(\frac{b-s+x}{2b}\right) -
  \left(b+s-x\right)\left(1+\frac{1}{\epsilon^2 b}\right)\right]} \\
  & \text{with} \quad x=\sqrt{(s+b)^2- 4 \epsilon^2 s b^2/(1+\epsilon^2 b)}.
\ee

\noindent Here $\epsilon$ is the relative systematic uncertainty of the 
background rate. In the special case of vanishing systematic uncertainty
$\epsilon \to 0$ these expressions simplify to

\begin{align}
 Z_\text{disc}^{\epsilon=0} = \sqrt{2[(s+b)\ln(1+s/b)-s]}, &&
 Z_\text{excl}^{\epsilon=0} = \sqrt{2[s-b\ln(1+s/b)]} 
\end{align}

\noindent In the limit of a  large number of background events, 
$b \gg s$, these expressions further simplify to the well known 
Gaussian approximations $Z_\text{disc} \approx s/\sqrt{b}$ and 
$Z_\text{excl} \approx s/\sqrt{s+b}$. In this work we choose a systematic 
uncertainty of $\epsilon=10\%$ for both the LHC and the  100 TeV collider.
We define regions with $Z_\text{disc} \geq 5$ as  discoverable regions, and 
regions with $Z_\text{excl} \leq 1.645$ as regions that can be excluded at  95\% CL.

\section{Simulation of Top-Tagging}
\label{sec:toptagging}
 
When an energetic top quark decays hadronically, its decay products are 
collimated and form a big jet, often called a \textit{fat jet}. The size of a 
top-initiated fat jet is given by $R \sim 2m_t/p_{T,t}$, which implies that 
only boosted top quarks with $p_{T}>250$ GeV will be able to form a fat jet 
of size $R<1.5$. While top-initiated fat jets show a characteristic 
substructure with subjets corresponding to the individual top decay products, 
such features are not present in QCD jets. Top-taggers are tools that analyze 
the fat jet's substructure to distinguish top-initiated from QCD initiated 
fat jets. Many ideas and techniques have been developed within the last year: 
QCD-based taggers like the HEPTopTagger ~\cite{Plehn:2010st,Plehn:2011sj,Kling:2012up}
or the Johns Hopkins Tagger~\cite{Kaplan:2008ie}, Event-shape based 
tagger like N-subjettiness~\cite{Thaler:2011gf} or template-overlap method 
based taggers like the TemplateTagger~\cite{Backovic:2012jk}. A (not so recent)
review about top tagging can be found in~\cite{Plehn:2011tg}. 

While most of the early taggers rely on only one analysis strategy, the more 
modern top taggers combine different approaches using machine learning tools. 
Examples include the HEPTopTagger Version-2~\cite{Kasieczka:2015jma}, the 
Deep-Top Tagger~\cite{Kasieczka:2017nvn} (focusing on low  $p_T$), and the Deep 
Neural Network Tagger~\cite{Pearkes:2017hku} (same idea, focusing on high  $p_T$).
A recent summary comparing modern top tagging approaches has been published 
by CMS~\cite{CMS:2016tvk}.

However, these  techniques are usually  computationally intensive, making 
them impractical for exploratory phenomenological studies such as this one. For this 
reason, we use a \emph{parametric} approach, implementing a \textsc{Delphes} 
top-tagging module inspired by the built-in \emph{b}-tagging module. We first 
reconstruct all fat jets with the size of $R=1.5$ using the Cambridge-Aachen 
algorithm~\cite{Dokshitzer:1997in} as implemented in \textsc{FastJet 3}~\cite{
Cacciari:2011ma}. We then assert that a  fat jet is top quark initiated if a 
parton-level top quark is found within a cone with a radius $R=0.8$ (we find 
that varying $R$ between 0.8 and 1.5 will not affect the results). 
Leptonically-decaying top quarks are rejected by vetoing fat jets with 
leptons in the jet cone. Once a fat jet is determined to be top-initiated, we 
apply a top-tagging efficiency $\epsilon_t$ for each of these fat jets. For 
QCD initiated fat jets, a misidentification rate $\epsilon_j$ is applied. 

\begin{figure}[t].
  \centering
  \includegraphics[width=0.48\textwidth]{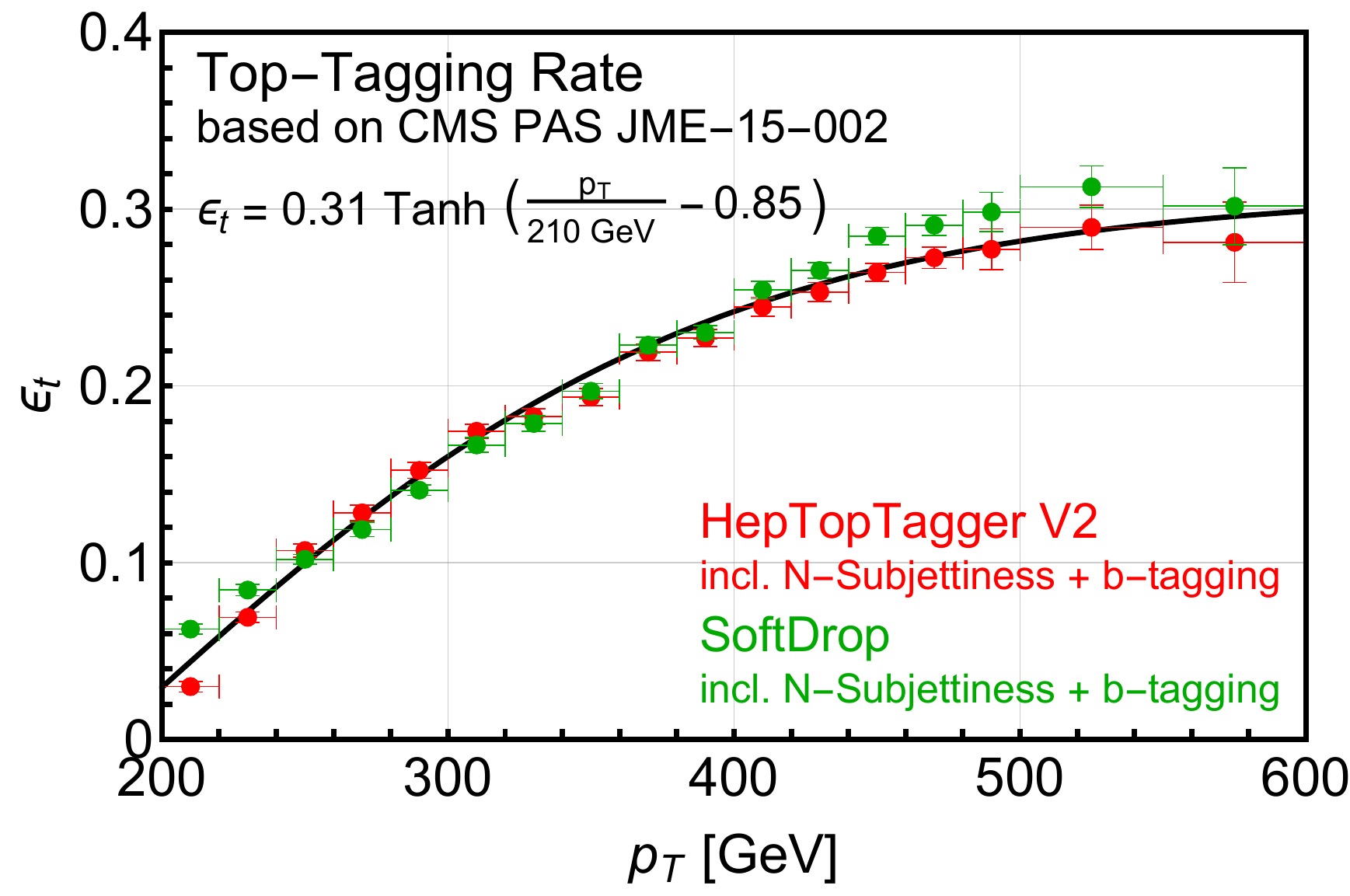}
  \includegraphics[width=0.48\textwidth]{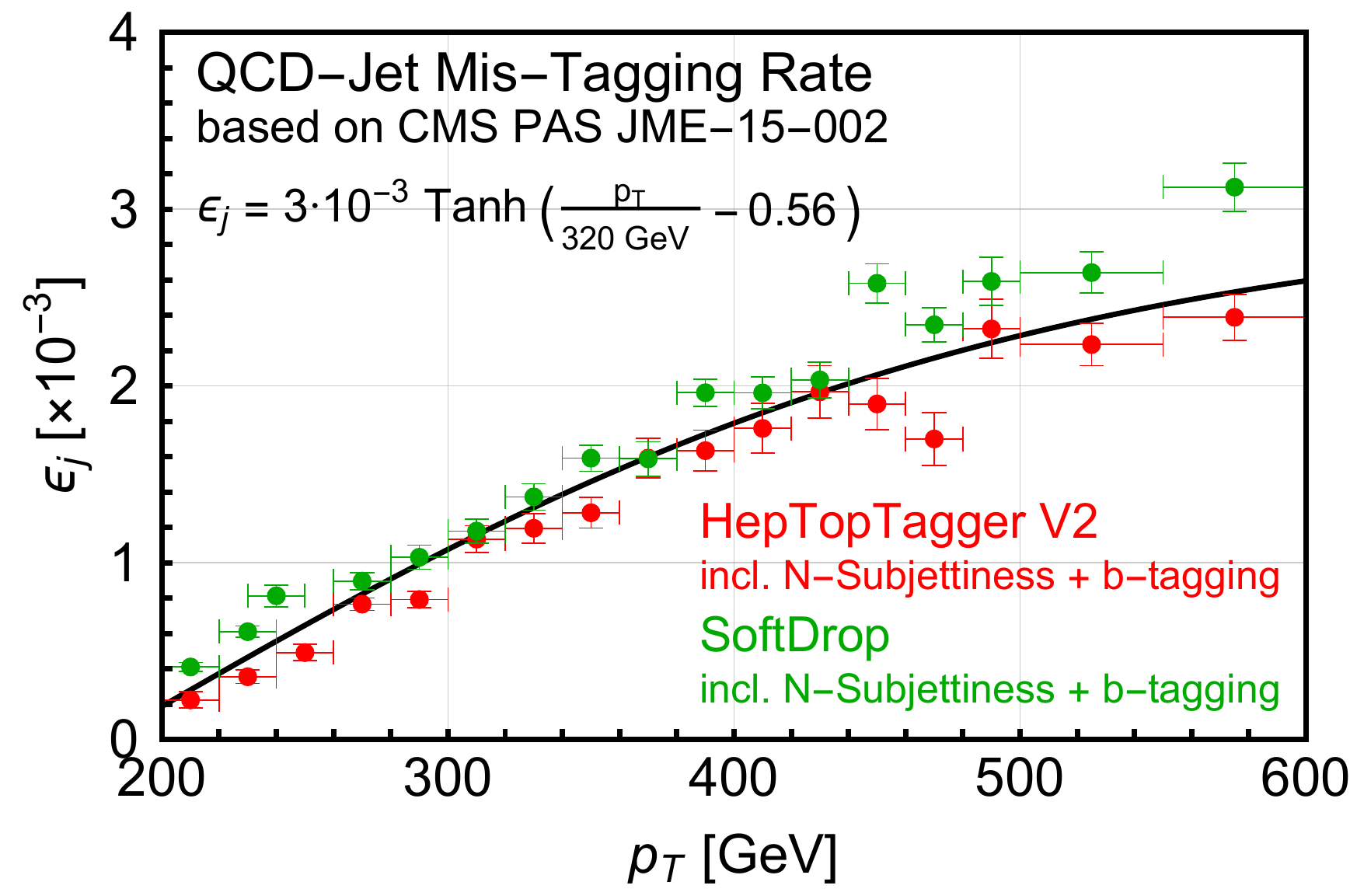}
  \caption{Top-tagging efficiencies (left) and QCD-jet mis-tagging rate (right) 
  for the HEPTopTagger (red) and SoftDrop (green) as adapted from the
  CMS study~\cite{CMS:2016tvk}.  The analytic parameterization used in this study is 
  shown as a   solid  black line.}
  \label{fig:top_tagging}
\end{figure}

In \figref{top_tagging} we show the top-tagging rate (left) and QCD-jet 
mis-tagging rate (right) as adapted from Fig. 10 in the CMS study~\cite{CMS:2016tvk}.
As representative examples we show the performance of the 
HEPTopTagger~V2~\cite{Kasieczka:2015jma} and SoftDrop~\cite{Larkoski:2014wba} 
in combination with groomed N-subjettiness and $b$-tagging. Both taggers have 
similar tagging and mis-tagging rates which are roughly independent of 
number of pile-up vertices. We parameterize their performance using an 
analytic form for top-tagging efficiency $\epsilon_t$ and  QCD-jet 
mis-identification rate $\epsilon_j$ and obtain
\be
 \epsilon_t = 0.31~\text{tanh}(p_T/210~\gev - 0.85) 
 \quad\text{and}\quad
 \epsilon_j = 0.003~\text{tanh}(p_T/320~\gev - 0.56). 
\ee

\bibliography{references}

\end{document}